\documentclass[]{imsart}
\usepackage{amsthm}
\usepackage{amsmath}
\usepackage{natbib}
\usepackage[colorlinks,citecolor=blue,urlcolor=blue,filecolor=blue,backref=page]{hyperref}
\usepackage{graphicx}

\usepackage{amssymb}
\usepackage{amsfonts}
\usepackage{multirow}
\usepackage{bm}
\usepackage{bbm}
\usepackage{color}
\usepackage{floatrow}
\usepackage{units}
\usepackage{url}
\usepackage{verbatim}
\usepackage{epstopdf}
\usepackage{booktabs,caption,fixltx2e}
\usepackage[flushleft]{threeparttable}
\usepackage{rotating}
\usepackage{multirow}
\usepackage{physics}
\usepackage{nameref}
\usepackage{placeins}
\usepackage{mathtools}
\usepackage[inline]{enumitem}

\startlocaldefs
\newtheorem{theorem}{Theorem}

 \newtheorem{proposition}[theorem]{Proposition}

\newcommand{\ind}{\stackrel{\mathrm{ind}}{\sim}}
\newcommand{\iid}{\stackrel{\mathrm{iid}}{\sim}}

\DeclarePairedDelimiterX{\infdivx}[2]{(}{)}{%
  #1\;\delimsize\|\;#2%
}

\makeatletter
\let\oldabs\abs
\def\abs{\@ifstar{\oldabs}{\oldabs*}}

\makeatletter 
\def\namedlabel#1#2{\begingroup
    #2%
    \def\@currentlabel{#2}%
    \phantomsection\label{#1}\endgroup
}
\makeatother
\endlocaldefs

\begin{document}


\begin{frontmatter}
\title{Pseudo Bayesian Estimation of One-way ANOVA Model in Complex Surveys}

\runtitle{One-way ANOVA in Complex Surveys}

\begin{aug}
\author{\fnms{Terrance D.} \snm{Savitsky}\thanksref{addr1}}
\and
\author{\fnms{Matthew R.} \snm{Williams}\thanksref{addr2}}
\and
\author{\fnms{Sanvesh} \snm{Srivastava}\thanksref{addr3}}

\runauthor{Savitsky, Williams, \and Srivastava}

\address[addr1]{U.S. Bureau of Labor Statistics, Office of Survey Methods Research
    \href{mailto:Savitsky.Terrance@bls.gov}{Savitsky.Terrance@bls.gov}
}

\address[addr2]{RTI International
    \href{mailto:mrwilliams@rti.org}{mrwilliams@rti.org}
}

\address[addr3]{Department of Statistics and Actuarial Science, The University of Iowa
    \href{mailto:sanvesh-srivastava@uiowa.edu}{sanvesh-srivastava@uiowa.edu}
}

\end{aug}

\begin{abstract}
We devise survey-weighted pseudo posterior distribution estimators under two-stage informative sampling of both primary clusters and secondary nested units for a one-way analysis of variance (ANOVA) population generating model as a simple canonical case where population model random effects are defined to be coincident with the primary clusters, for example student performance based on a survey of schools and students such as the 2000 OECD Programme for International Student Assessment (PISA). We consider estimation on an observed informative sample under both an augmented pseudo likelihood that co-samples the random effects, as well as an integrated likelihood that marginalizes out the random effects from the survey-weighted augmented pseudo likelihood.  This paper includes a theoretical exposition that enumerates easily verified conditions for which estimation under the augmented pseudo posterior is guaranteed to be consistent at the true generating parameters. We reveal in simulation that both approaches produce asymptotically unbiased estimation of the generating hyperparameters for the random effects when a key condition on the sum of within cluster weighted residuals is met.  We present a comparison with two frequentist alternatives, an expectation-maximization approach and a composite likelihood method that requires pairwise sampling weights.
\end{abstract}



\end{frontmatter}

\section{Introduction}

The current literature for Bayesian methods has partially addressed population model estimation of survey data through the use of marginal survey sampling weights to obtain consistent estimates of fixed effects or top level global parameters estimated on survey data under an informative, complex sampling design. An informative sampling design constructs the known sampling inclusion probabilities to be correlated with a response variable of interest (e.g., student performance). This correlation produces observed samples with a different distribution for the response variable from that of the underlying population of focus for analysis.  The marginal sampling weights are used to formulate a pseudo likelihood that exponentiates each unit likelihood contribution by its sampling weight, treated as fixed, to produce an approximate likelihood for the population, estimated on the observed sample taken from that population. The pseudo likelihood re-balances information in the observed sample to approximate that in the population.  A pseudo posterior distribution for population model parameters results from convolving the pseudo likelihood with prior distributions on the population model parameters.  The use of the pseudo posterior may be situated in the more general class of approximate or composite likelihoods \citep{2009arXiv0911.5357R}.   \citet{williams2020} demonstrate consistency of the resulting pseudo posterior distribution under mild conditions that define a class of admitted sampling designs that allow for unattenuated dependence in size-restricted clusters.  The consistency result restricts the complexity (of the parameterization) of the underlying population generating model.

Our focus in this paper is on mixed effects modeling for estimation of random effects and their generating parameters under a two-stage sampling design of clusters, followed by units nested within clusters. Both the clusters and units within cluster may be sampled informatively from the population.  Our inferential interest lies in estimation of cluster-indexed random effects and their generating variances. The Organisation for Economic Co-operation and Development (OECD) Programme for International Student Assessment (PISA) 2000 is a collection of surveys of schools and students conducted across 32 countries.
The PISA 2000 employed a two-stage school-based design \citep{organisation2000database}, with schools being selected with probability proportional to enrollment size and students within each school selected with equal probability. In particular, larger schools are more likely to be selected and may have systematically different performance outcomes than smaller schools. While the number of students within each school may vary, the target number of students is mostly fixed. Thus the marginal selection probabilities of individual students varies across school and we would expect that the intercept and student variance contribution may be estimated with error if the sample design is ignored.  The classic ANOVA model is a decomposition of variance. For PISA 2000 we seek to decompose the variance of student performance into school-level and student-level contributions.

The consistency result of \citet{williams2020} could be readily interpreted to incorporate any hierarchical Bayesian model used, in practice, when the inferential focus is on parameters that appear in the observed data likelihood.  \citet{williams2020} focus only on models parameterized solely by fixed effects.  Yet, it is routine in Bayesian modeling to employ one or more sets of random effects under prior formulations designed to capture complex covariance structures.  The inferential focus is on the generating hyperparameters of the random effects, which don't appear in the observed data likelihood.  Hierarchical specifications make such population models readily estimable.  The survey statistics literature suggests that estimating the hyperparameters of the prior distribution for the random effects are still potentially estimated with bias \citep{rh:1998}. The possibility for survey-induced bias in the estimation of random effects severely limits the applicability of the full suite of Bayesian models to complex social and economic data. \citet{SavitskyWilliams+2022+901+928} propose a Bayesian framework that  provides a principled solution. They demonstrate that weighting both the likelihood and the prior distribution for the random effects to produce an augmented pseudo likelihood for the data and random effects performs well and they demonstrate consistency of the pseudo posterior estimator with respect to the joint distribution of population generation and the taking of samples from the finite population. 
Applying their framework to the specific case of a one-way ANOVA (with Gaussian components denoted by $\mathcal{N}()$) leads to an \emph{augmented} pseudo likelihood, $\mathcal{N}\left(y_{j k}\middle\vert \mu + a_{k},\tau_{\epsilon}^{-1}\right)^{w_{jk}} \times  \mathcal{N}\left(a_{k}\middle\vert 0,\tau_{a}^{-1}\right)^{w_{k}}$, where $k$ indexes cluster and $j$ indexes a unit nested within cluster $k$.  The observed response variable for sampled unit $(j,k)$ is denoted by $y_{jk}$ and latent random effect for cluster $k$ is denoted by $a_{k}$.  The cluster $k$ marginal sampling weight is denoted by $w_{k}$ and the unit $(j,k)$ marginal sampling weight is denoted by $w_{jk}$.  This approach extends the pseudo likelihood approach from the fixed effects model specification to the estimation of mixed effects where clusters are informatively sampled.  The inferential focus is on global generating parameters, $(\mu,\tau_{\epsilon}^{-1},\tau_{a}^{-1})$.

We proceed to introduce Bayesian mixed effects models, estimated from survey data, to provide context for our specifications of conditions that guarantee a consistent estimation of global generating parameters under the one-way ANOVA model. 

\section{Bayesian Mixed Effects Estimation}

We are interested in model-based Bayesian inference on the fixed and random effect parameters in multistage designs.
Let $U_c = \{1,\ldots,M\}$ and $h \in U_c$ index clusters of units of size $N_{h}$ in the unobserved population from which we will take a sample.
Let $y_{\ell h}$ denote the population response from the unit $\ell$ nested in the cluster $h$, $a_h$ denote the population random effect specific to the cluster $h$, and $\theta$ and $\phi$ denote the fixed effect and random effects parameters for generating $y_{\ell h}$ and $a_h$, respectively, where $h \in U_c$ and $\ell \in \{1,\ldots, N_h\}$. Let $f(\cdot \mid a_h, \theta)$ and $f(\cdot \mid \phi)$ denote the conditional densities for the finite population responses and their random effects, respectively. Then our pseudo posterior estimation approach begins with a complete joint model for the finite population, $U$, as if the random effects, $\bm{a}_{U} = \{a_{h}\}$, were directly observed just like the response, $\bm{y}_{U} = \{y_{ \ell h}\}$,
\begin{equation}\label{eq:popmodel}
  f_{U}\left(\theta, \phi \vert  \bm{y}_{U}, \bm{a}_{U} \right) \propto
  \left[  \mathop{\prod}_{h = 1}^{M}
    \left(\mathop{\prod}_{\ell= 1}^{N_{h}} f\left(y_{ \ell h}\middle\vert a_{h}, \theta \right)\right)
    f\left(a_{h}\middle\vert \phi \right)  \right]f(\theta)f(\phi),
\end{equation}
where $f(\theta)$ and $f(\phi)$ are densities of the prior distributions of $\theta$ and $\phi$, respectively. This is analogous to specifying a `complete data' likelihood when constructing an EM algorithm \citep{10.2307/2984875}.

The sampling design distribution, which governs all possible samples of the population $U$, is assumed to be known, conditional on the realized values of the population response $\bm{y}_{U}$ and cluster level random effects $\bm{a}_{U}$. Let $\mathbb{P}^{\pi}$ denote the sampling design distribution which is specified on a vector of marginal inclusion indicators for the population of clusters, $(\delta_{1},\ldots,\delta_{M}) \in \{0, 1\}^M$, for the clusters $h \in \{1,\ldots, M \}$ and conditional inclusion indicators given a cluster $h$, $(\delta_{1\vert h},\ldots,\delta_{N_{h} \vert h})$, for the $N_h$ units nested in cluster $h$. Let $S_{c} = \{h \in (1,\ldots,M): \delta_{h}=1  \}\subseteq U_{c}$ denote a random sample of clusters with $\vert S_{c} \vert = m$.
Fix a sampled cluster $k \in S_{c}$ and let $S_{k} = \{\ell \in (1,\ldots,N_{k}): \delta_{\ell \vert k} = 1\}$ with $\vert S_{k} \vert = n_{k}$, denote a random sample of nested units. $\mathbb{P}^{\pi}$ governs the selection of random sample, $S$, of nested units across clusters, where $\vert S \vert \equiv n = \sum_{k=1}^{m}n_{k}$. $\mathbb{P}^{\pi}$ is expressed through the specifications of marginal inclusion probabilities for clusters, $\pi_{h} \equiv P(\delta_{h} = 1\vert \{y_{\ell h}\}_{\ell}, a_{h})$ for all $h \in U_{c}$ and conditional inclusion probabilities, $\pi_{\ell\vert g} = P(\delta_{\ell\vert g} = 1 \vert \{y_{\ell g}\}_{\ell})$, where $g \in S_{c}$ denotes a cluster included in a random sample, and where the notation $\{\cdot_{ab}\}_{a}$ is used to indicate a subset of $\{\cdot_{ab}\}$ for which the $b$ index is fixed.
The conditioning on $\delta_{g} = 1$ in the statement of $\pi_{\ell\vert g}$ is implied in definition of $\delta_{\ell\vert g}$. The conditioning of each $\pi_{h}$ on response values within cluster, $\{y_{\ell h}\}_{\ell}$, implies that $\delta_{h} \perp \delta_{h^{'}}$ for every  $h \neq h'\in (1,\ldots,M) $.
We utilize $\mathbbm{1}(j \in S_{k})$ to denote unit, $j \in (1,\ldots,n_{k})$ in sampled cluster $k$ where $\delta_{j\vert k} = 1$, where we have re-numbered each of selected clusters and units selected from $U$ to be in sequence in an abuse of notation and without loss of generality.  Similarly, $\mathbbm{1}(k\in S_{c}) \equiv (\delta_{k}=1)$ and, finally, $\mathbbm{1}((j,k) \in S)\equiv \delta_{jk} \equiv (\delta_{k} \times \delta_{j\vert k}) = 1$. To summarize: we use $(l,h)$ when referring to every individual and cluster in the population and we use $(j,k)$ when referring to individuals and their corresponding clusters that are selected in a realized sample.

We construct sampling weights used for estimation from an \emph{observed} sample as $w_{k} = 1/\pi_{k}$ and $w_{j\vert k} = 1/\pi_{j\vert k}$ such that their composition, $w_{jk}\equiv w_{k} \times w_{j\vert k}$ defines the marginal sampling weight for unit, $(j,k)\in S$.
The sampling-weighted model approximation to Equation \ref{eq:popmodel} based on the observed sample $\bm{y}_{S} = \{y_{ j k}\}$ and $\bm{a}_{S} = \{a_{k}\}$ is
\begin{equation}\label{eq:sampcompletemodel}
  f^{\pi}\left(\theta, \phi \vert  \mathbf{y}_{S}, \mathbf{a}_{S}\right) \propto \left[\mathop{\prod}_{k =1}^m\left(\mathop{\prod}_{j =1}^{n_k} f\left(y_{ j k}\middle\vert a_{k}, \theta \right)^{w_{k j}}\right)
    f\left(a_{k}\middle\vert \phi \right)^{w_k} \right]f(\theta)f(\phi)
,
\end{equation}
and a sampling-weighted version of the observed model is
\begin{equation}\label{eq:sampobsmodel}
  f^{\pi}\left(\theta, \phi \vert  \mathbf{y}_{S} \right) \propto
\left[\mathop{\int}_{\mathbf{a} \in \mathcal{A}} \left\{ \mathop{\prod}_{k =1}^m
\left(\mathop{\prod}_{j = 1}^{n_k} f\left(y_{ j k}\middle\vert a_{k}, \theta \right)^{w_{k j}}\right)
f\left(a_{k}\middle\vert \phi \right)^{w_k} \right\} d \mathbf{a} \right]f(\theta)f(\phi);
\end{equation}
see \citet{SavitskyWilliams+2022+901+928} for greater details. This observed model is analogous to an `incomplete data' likelihood for an EM algorithm \citep{10.2307/2984875}

We are concerned with achieving asymptotically unbiased inference for $(\theta,\phi)$ estimated on the observed sample taken under an informative survey sampling design in the one-way ANOVA model. Specifically, the simulation study comparison and theoretical results for the augmented pseudo posterior in Equation \ref{eq:sampcompletemodel} and the integrated pseudo posterior in Equation \ref{eq:sampobsmodel} are developed for the canonical case of a one-way ANOVA mixed effects population generation and estimation. The  population model for the one-way ANOVA (as though it were fully observed) is specified as,
\begin{equation}\label{eq:estmod}
\begin{array}{rl}
  f\left(y_{\ell h}\middle\vert a_{h},\theta\right) \ \equiv & \mathcal{N}\left(\mu + a_{h},\tau_{\epsilon}^{-1}\right)\\
f\left(a_{h}\middle\vert\phi\right) \ \equiv & \mathcal{N}\left(0,\tau_{a}^{-1}\right)\\
f(\mu) \ = & \mathcal{U}\left(-\infty,\infty\right)\\
f\left(\tau_{a}^{-1}\middle\vert \alpha_{1},\beta_{1}\right) \ \equiv & \mathcal{IG}(\alpha_{1},\beta_{1})\\
f\left(\tau_{\epsilon}^{-1}\middle\vert \alpha_{2},\beta_{2}\right) \ \equiv & \mathcal{IG}(\alpha_{2},\beta_{2}),
\end{array}
\end{equation}
with $\theta \equiv (\mu,\tau_{\epsilon}^{-1})$ and $\phi \equiv \tau_{a}^{-1}$ and $\mathcal{N}(\cdot,\cdot)$, $\mathcal{U}(\cdot,\cdot)$ and $\mathcal{IG}(\cdot,\cdot)$ represent normal, uniform and inverse gamma distributions, respectively. For simplicity of exposition we use an improper uniform prior for $\mu$.  We validated by computation that the joint distribution for the model of Equation~\ref{eq:estmod} is a valid (integrable) distribution function and provide insights on this computation in Section~\ref{sec:ilike}.  

In section {\ref{sec:consistency}} we specify assumptions that guarantee consistency of the full conditional posterior distributions (that formulate the joint distribution) for (global) parameters $(\mu,\tau_{a}^{-1},\tau_{\epsilon}^{-1})$.

The simulation study of Section~\ref{sec:sims1} demonstrates that conducting Bayesian estimation under the augmented pseudo posterior distribution (by co-sampling $\{a_{k}\}$)  of Equation~\ref{eq:sampcompletemodel} is equivalent to sampling the marginalized pseudo posterior distribution of Equation~\ref{eq:sampobsmodel} and both produce asymptotically unbiased estimation of $(\theta,\phi)$ if a condition on balanced within cluster weighted residuals is met.  The augmented pseudo posterior employs a sampling weighted pseudo likelihood that accomplishes estimation adjustments for the informativeness of sampling both clusters, $k$, and nested units, $j$. For the one-way ANOVA, the augmented pseudo posterior admits proper full conditional pseudo posterior distributions which will be used in Section \ref{sec:consistency} to demonstrate frequentist consistency.

We derive the integrated likelihood of Equation~\ref{eq:sampobsmodel} in Section~\ref{sec:ilike} for the one-way ANOVA model and show that it includes the product of sampling weighted normal distribution kernels that function in the same fashion as does the augmented likelihood to adjust estimation on the observed sample. While the augmented likelihood framework included in Equation~\ref{eq:sampcompletemodel} produces conjugate full conditional pseudo posterior distributions that are easily sampled in a Gibbs scan, there are many posterior sampling algorithms that may be employed.  In the sequel, we utilize the implementation of the Hamiltonian Monte Carlo (HMC) sampler implemented in Stan \citep{stan:2015}.  HMC is a specific formulation of the Metropolis algorithm that partially suppresses its random walk properties to achieve more rapid convergence on the target distribution.  When the resulting joint posterior distribution is a valid, integrable distribution function, both HMC and Gibbs sampling are guaranteed to achieve the target pseudo posterior distribution by the ergodic theorem and regularity conditions imposed on the class of allowable sampling designs that bound the sampling weights.  Our use of {HMC} to conduct pseudo posterior estimation is fortuitous because it does not require conjugacy as all model parameters are jointly sampled in each iteration sweep.  Unlike in the case of the augmented pseudo likelihood, the integrated pseudo likelihood contained in Equation~\ref{eq:sampobsmodel} does \emph{not} produce closed-form full conditional pseudo posterior distributions, but such is no problem for sampling under HMC. In addition, we utilize Equation~\ref{eq:sampobsmodel} to solve for the maximum a-posteriori point estimates via optimization. The analytic integration of the cluster level random effects greatly reduces the dimension and thus standard optimization methods on the remaining small number of parameters are more likely to succeed.

In contrast, \citet{pfeffmix:1998} and \citet{rh:1998} specify the following integrated likelihood under frequentist estimation for an observed sample where units are nested within clusters,
\begin{equation}\label{eq:wklike}
\ell^{\pi}(\theta,\phi) = \mathop{\sum}_{k=1}^{m} w_{k} {\ell_{k}^{\pi}(\theta,\phi)} ,
\end{equation}
for {$\ell_{k}^{\pi}(\theta,\phi) = \log L_{k}^{\pi}(\theta,\phi)$} and
\begin{align}\label{eq:frequnitlike}
  {L_{k}^{\pi}(\theta,\phi) = \mathop{\int}_{a_{k} \in \mathcal{A}}\exp\left[\mathop{\sum}_{j =1}^{n_k} w_{j \vert k}\ell \left(y_{ j k} \middle\vert a_{k}, \theta \right)\right]f\left(a_{k}\middle\vert \phi \right)da_{k},}
\end{align}
which will \emph{not}, generally, be design unbiased for the population likelihood because the unit level conditional weights, $\{w_{j\vert k}\}$, are nested inside an exponential function; therefore, replacing $w_{j\vert k}$ with $\delta_{\ell \vert h}/\pi_{\ell \vert h}$ inside the exponential and summing over the population clusters and nested units will not produce separable sampling design terms that each integrate to $1$ with respect to $\mathbb{P}^{\pi}$ conditioned on the generated population \citep{yi:2016}.

It is interesting to note, however, that the pseudo likelihood of \citet{rh:1998} collapses onto Equation~\ref{eq:sampcompletemodel} under Bayesian estimation employing data augmentation, as we see with,
\begin{align}
  L^{\pi}_{{k}}\left(\theta,\phi \middle\vert \mathbf{y}_{S},\mathbf{a}_{S}\right) &= \exp\left[ \mathop{\sum}_{j =1}^{n_k} w_{j\vert k} \ell\left(y_{j k}\middle\vert a_{k},\theta\right) \right]f\left(a_{k}\middle\vert \phi\right)\\
\log L^{\pi}_{{k}}\left(\theta,\phi\middle\vert \mathbf{y}_{S},\mathbf{a}_{S}\right) &= \left[ \mathop{\sum}_{j =1}^{n_k} w_{j\vert k} \ell\left(y_{j k}\middle\vert a_{k},\theta\right) \right] + \ell\left(a_{k}\middle\vert \phi\right)\\
\ell^{\pi}\left(\theta,\phi\middle\vert \mathbf{y}_{S},\mathbf{a}_{S}\right) &= \mathop{\sum}_{k=1}^{m} w_{k}\left[ \mathop{\sum}_{j =1}^{n_k} w_{j\vert k} \ell\left(y_{j k}\middle\vert a_{k},\theta\right) + \ell\left(a_{k}\middle\vert \phi\right)\right]\\
&= \mathop{\sum}_{k=1}^{m}\mathop{\sum}_{j =1}^{n_k} w_{kj} \ell\left(y_{j k}\middle\vert a_{k},\theta \right) + \mathop{\sum}_{k=1}^{m} w_{k}\ell\left(a_{k}\middle\vert \phi\right).
\end{align}

\citet{pfeffmix:1998} informally discuss that their estimator of Equation~\ref{eq:frequnitlike} is nevertheless consistent for estimation of $\phi$ if both the number of sampled clusters, $m$ and the number of within cluster sampled units, $n_{k}$, \emph{both} limit to $\infty$.  By contrast, we will formally show in the sequel that consistency is achieved for $\phi$ of Equation~\ref{eq:sampcompletemodel} without requiring $n_{k}$ to limit to $\infty$, which we believe accords with practical sampling design settings; for example, if one is sampling households in a geographically-indexed primary sampling unit (PSU), one would increase the number of PSUs sampled to increase estimation power for a domain-indexed (e.g., a state or county) statistic, recognizing that the number of households within each PSU remains relatively constant.  We are able to remove the condition on $n_{k}$ because our approach does not require consistency for each random effect, $a_{k}$, to achieve consistency of $\phi$ but rather requires consistency for the sampling weighted \emph{mean} of the random effects. The sampling weighted least squares estimator of \citet{pfeffmix:1998} does require consistency of each $a_{k}$. Perhaps it might be possible to achieve consistency of their estimator without this condition on $a_{k}$ if one uses a similar approach as do we for our Bayesian estimator in Section~\ref{sec:consistency}.

Taking a marginal integration approach, \citet{slud_2019} assesses consistency of MLE estimation under both Equations~\ref{eq:frequnitlike} and \ref{eq:sampobsmodel} for a one-way ANOVA population generating and estimation models under two-stage informative sampling of clusters and of units within clusters.  They demonstrate that Equation~\ref{eq:frequnitlike} is not consistent for MLE estimation under informative sampling of clusters and non-informative sampling of nested units when assuming the within cluster sample size, $n_{k}$, is bounded from above (rather than limiting to $\infty$), which is coherent with \citet{pfeffmix:1998}. Their simulation study for the one-way ANOVA model reveals that $\phi$ is estimated with asymptotic bias due to the failure in application of the cluster-indexed weight, $w_{k}$, \emph{after} integrating out each random effect, $a_{k}$, to correct for bias. \citet{slud_2019} also shows that the marginal pseudo likelihood contained in Equation~\ref{eq:sampobsmodel}, which we use for our integrated pseudo posterior estimation, will also not provide consistent MLE estimation for the one-way ANOVA in the particular case the cluster sampling fraction, $m/M$  limits to $0$ as $M \rightarrow \infty$ in the case that the cluster sampling weights, $w_{k}$, also limit to $\infty$ for some $k \in (1,\ldots,m)$ or in the separate case of $\rho_{k} = N_{k}\tau_{a}^{-1}/(N_{k}\tau_{a}^{-1} + \tau_{\epsilon}^{-1}) \ll 1$. In the latter case, if $\tau_{\epsilon}$ and $\tau_{a}$ are of similar magnitudes, then this implies that $N_k$ has to be very small.

In contrast, we show in Section~\ref{sec:consistency} that the full conditional pseudo posterior distribution for $\phi$ under the augmented pseudo posterior of Equation~\ref{eq:sampcompletemodel} for the one-way ANOVA model is guaranteed to be consistent under the requirement that $m/M \rightarrow f_{c} > 0$ as $M \rightarrow \infty$; in other words, our estimator would \emph{not} be consistent if the group sampling fraction limits to $0$ (\emph{and} $w_{k}$ limits to $\infty$ for some $k \in (1,\ldots,m)$), so that our result is coherent with \citet{slud_2019}.  Yet, our condition for a non-zero asymptotic sampling fraction is not restrictive.  The assumption of an asymptotically $0$ sampling fraction is often done to simplify variance estimation for a survey estimator of interest \citep{10.2307/1403631}.

More importantly, our consistency result in the sequel includes a condition that requires the expectation of the within cluster weighted sample residuals to be balanced (at $0$) for all clusters in the population.  This condition defines a class of sampling designs for which consistency is  achieved (e.g., balanced sampling designs) or nearly achieved (e.g. weakly unbalanced sampling designs) and, by contrast, a class of sampling designs under which consistency would not be expected to be achieved (e.g., highly unbalanced sampling designs).  

\section{Consistency of Bayesian Estimators}\label{sec:consistency}

\subsection{Preliminaries}

Consider the setup for the true population generating model. Let $Y_{1 1}, \ldots, Y_{N_1 1}, \ldots, Y_{1 M}, \ldots, Y_{N_M M} \sim \mathbb{P}_{\lambda_{0}}$ be sequence
of conditionally independent but non-identically distributed random variables defined for a population of clusters $U_c$, where the size of $U_c$ is $M$. For a cluster $h \in \{1, \ldots, M\}$, $U_{h}$ denotes the population of units $N_{h}$ nested within cluster $h$, where  $N_h$ is the size of $U_h$. We use the index $\displaystyle(\ell,h)$ to denote the $\ell$th unit in the cluster $h$, and the total number of $(\ell, h)$ units in the population equals $N = \sum_{h=1}^M N_h$. The true population generating model is a one-way ANOVA, specified as
\begin{equation}\label{poptruth}
\begin{array}{rl}
Y_{\ell h} = \mu_{0} + a_{h0} + \epsilon_{\ell h 0}, \quad
a_{h0} \iid \mathcal{N}(0,\tau_{a 0}^{-1}),\quad
\epsilon_{\ell h 0} &\iid \mathcal{N}(0,\tau_{\epsilon 0}^{-1})
\end{array}
\end{equation}
for $(\ell, h) \in \{1, \ldots, N\}$, where $\mu_0$ is the intercept, $\{a_{h0}\}$ is the array of the random effects, $\epsilon_{h\ell0}$ are the idiosyncratic errors, $\mathcal{N}$ indicates a Normal distribution, and $\lambda_{0} = (\{a_{h0}\},\mu_{0},\tau_{a0}^{-1},\tau_{\epsilon 0}^{-1}) \in \Lambda$ are the true generating parameters.

Under the truth of Equation~\ref{poptruth}, we specify an estimation model for the population (as if the population, $\{y_{\ell h}\}$ were fully observed under latent random effects $\{a_{h}\}$) with,
\begin{equation}\label{popmodel}
\begin{array}{rl}
y_{\ell h}\vert \mu, a_h, \tau_{\epsilon}^{-1}  &\ind \mathcal{N}\left(y_{\ell h}\middle\vert \mu + a_{h},\tau_{\epsilon}^{-1}\right)\\
a_{h} \vert \tau_{a}^{-1} &\ind \mathcal{N}\left(a_{h}\middle\vert 0,\tau_{a}^{-1}\right)\\
\mu   &\sim \mathcal{U}(-\infty, \infty)\\
\tau_{a}^{-1} \vert \alpha_{1}, \beta_{1} &\sim \mathcal{IG}(\alpha_{1},\beta_{1})\\
\tau_{\epsilon}^{-1} \vert \alpha_{2}, \beta_{2} &\sim \mathcal{IG}(\alpha_{2},\beta_{2}),
\end{array}
\end{equation}
where $\mathcal{U}$ is a uniform distribution, and $\mathcal{IG}$ an inverse gamma distribution. Collect estimation parameters, $\lambda = (\{a_{h}\},\mu,\tau_{a}^{-1},\tau_{\epsilon}^{-1}) \in \Lambda$, an $M + 3$ space of parameters measurable with respect to prior distribution, $\Pi$, (on the space, $\Lambda$) with above-specified densities.  Our inferential interest is in global parameters, $\theta = (\mu, \tau_{a}^{-1},\tau_{\epsilon}^{-1}) \in \Theta$ under the population model of Equation~\ref{popmodel}.

Together $\mathbb{P}_{\lambda_{0}},\mathbb{P}^{\pi}$ index the joint distribution over the population generation and the taking of a random sample from the population. In the sequel, we use $\mathbb{P}_{\lambda_{0}},\mathbb{P}^{\pi}$ to compute expectations of a (pseudo) Bayesian estimator to achieve an $L_{1}$ frequentist consistency result. We define the associated rate of convergence notation, $a = \mathcal{O}(b)$, to denote $ |a| \le C |b|$ for a constant $C > 0$.

\subsection{Model specifications}

Our focus is inference on the true generating parameters $\theta_{0}= \{\mu_0, \tau_{a0}^{-1}, \tau_{\epsilon 0}^{-1}\}\in \Theta$ by specifying the augmented parameters $\lambda\in\Lambda$ under prior distribution $\Pi$ specified for the population generating model. Our approach generates intermediate estimates for the random effects $\{a_{h}\}$, but we are not interested in them individually. In fact, we expect these individual random effect estimates to be biased for small within cluster samples due to shrinkage; however, globally across all random effects, this produces a James-Stein like estimator that would achieve smaller total loss (e.g., mean square error $M^{-1}\sum_{h} (a_{h} - a_{h0})^2$) than a fixed effects model.

We approximate the target population estimation model of Equation~\ref{popmodel} for a random sample in $S$ with
\begin{equation}\label{randsampmodel}
  \begin{array}{rl}
    Y_{\ell h}\vert \mu, a_h, \tau_{\epsilon}^{-1}, w_{\ell h}  &\ind \mathcal{N}\left(y_{\ell h}\middle\vert \mu + a_{h},\tau_{\epsilon}^{-1}\right)^{\frac{\delta_{\ell h}}{\pi_{\ell h}}}\\
    a_{h} \vert \tau_{a}, w_{h} &\ind \mathcal{N}\left(a_{h}\middle\vert 0,\tau_{a}^{-1}\right)^{\frac{\delta_{h}}{\pi_{h}}}\\
    \mu   &\sim \mathcal{U}(-\infty, \infty)\\
    \tau_{a}^{-1} \vert \alpha_{1}, \beta_{1} &\sim \mathcal{IG}(\alpha_{1},\beta_{1})\\
    \tau_{\epsilon}^{-1} \vert \alpha_{2}, \beta_{2} &\sim \mathcal{IG}(\alpha_{2},\beta_{2}),
  \end{array}
\end{equation}
where $h\in (1,\ldots,M)$ and $\ell \in (1,\ldots,N_{h})$ and
the notation $\mathcal{N}(x \vert \mu, \sigma^2)^{e}$ indicates a distribution whose density is proportional to a normal density function raised to an exponent: $\phi(x,\mu, \sigma^2)^{e}$. 
The sampling weighted pseudo likelihood for $Y_{\ell h}$ is not a generative likelihood, but a noisy approximation for the likelihood for the unobserved population. Similarly, the sampling weighted pseudo prior for $a_{h}$ is also an approximation to the population prior distribution for the unobserved random effects. We use this formulation to develop a pseudo posterior distribution for $\lambda$, which we subsequently use to demonstrate frequentist consistency with respect to the joint distribution, $\mathbb{P}_{\lambda_{0}},\mathbb{P}^{\pi}$, governing the generation of a population and the taking of a sample from that population.
The model in Equation~\ref{randsampmodel} defines a pseudo posterior distribution for $\theta$, $\Pi\left(\theta \middle \vert \{y_{\ell h}\}, \{a_{h}\},\{\delta_{\ell h}\},\{\pi_{\ell h}\}\right)$, which is our Bayesian estimator.  From this distribution, we subsequently derive summary measures (e.g., the first two moments) and compute expectations of these summary measures with respect to the joint distribution $\mathbb{P}_{\lambda_{0}},\mathbb{P}^{\pi}$ to demonstrate $L_{1}$ contraction of the entire pseudo posterior distribution $\Pi\left(\theta \middle \vert \{y_{\ell h}\}, \{a_{h}\},\{\delta_{\ell h}\},\{\pi_{\ell h}\}\right)$ on $\theta_{0}$ with respect to the $\mathbb{P}_{\lambda_{0}},\mathbb{P}^{\pi}$ distribution.

Our estimating model on an \emph{observed} sample $\{y_{jk},\delta_{jk} = 1, \delta_{k} = 1\}_{jk}$ is specified by,
\begin{equation}\label{obssampmodel}
  \begin{array}{rl}
    y_{j k}\vert \mu, a_k, \tau_{\epsilon}^{-1}, w_{jk}  &\ind \mathcal{N}\left(y_{j k}\middle\vert \mu + a_{k},\tau_{\epsilon}^{-1}\right)^{w_{jk}}\\
    a_{k} \vert \tau_{a}^{-1}, w_{k} &\ind \mathcal{N}\left(a_{k}\middle\vert 0,\tau_{a}^{-1}\right)^{w_{k}}\\
    \mu   &\sim \mathcal{U}(-\infty, \infty)\\
    \tau_{a}^{-1} \vert \alpha_{1}, \beta_{1} &\sim \mathcal{IG}(\alpha_{1},\beta_{1})\\
    \tau_{\epsilon}^{-1} \vert \alpha_{2}, \beta_{2} &\sim \mathcal{IG}(\alpha_{2},\beta_{2}),
  \end{array}
\end{equation}
While neither the pseudo likelihood or pseudo prior are proper distributions, the augmented pseudo likelihood normalizes to a proper normal distribution. (See Equation 15c and the follow-on discussion in \citet{2015arXiv150707050S} that demonstrates that the augmented pseudo likelihood is normalizable to a proper normal distribution with precision updated by the sampling weights). We have subsequently verified that the resulting joint pseudo posterior distribution, $\Pi(\lambda\vert \{y_{jk}\}, \{w_{jk}\}, \{w_{k}\})$ is integrable. See Section~\ref{sec:ilike} for further discussion and insight into the integrability of the joint pseudo posterior distribution. Even though the model of Equation~\ref{obssampmodel} uses an improper prior $\mu \propto 1$, we straightforwardly validated that the joint distribution for this hierarchical probability is a valid (integrable) distribution function such that the full conditionals will contract onto the correct target joint distribution.  Insight into the computation for the joint distribution may be gleaned from the form  of the integrated likelihood that marginalizes the $\{a_{k}\}_{k}$ in Section~\ref{sec:ilike}. The integrated pseudo likelihood is a function of global parameters, $(\mu,\tau_{a}^{-1},\tau{\epsilon}^{-1})$, and the integrated pseudo posterior after incorporating prior $\mu \propto 1$ is integrable. The use of an improper prior for the mean parameter is typical in order to make the Bayesian probability model produce the same posterior mean as the frequentist MLE by avoiding shrinking \citep{gelmanbda04,RePEc:bla:jorssa:v:176:y:2013:i:3:p:795-808}.  

We deconstruct the joint pseudo posterior distribution under a Gibbs scan,
\begin{equation}\label{gibbsscan}
\begin{array}{rl}
&\left(a_{k} \middle\vert \tau_{a}^{-1}, \tau_{\epsilon}^{-1}, \mu,\{y_{jk}\}_{j},\{w_{jk}\}_{j},w_{k}\right), \quad k = 1,\ldots,m\\
&\left(\mu \middle\vert \{a_k\}, \tau_{\epsilon}, \{y_{jk}\}\right) \\
&\left(\tau_{a}^{-1}\middle\vert \{a_k\}, \alpha_{1}, \beta_{1}\right)\\
&\left(\tau_{\epsilon}^{-1}\middle\vert \{a_{k}\},\mu, \{y_{jk}\}, \{w_{jk}\}, \alpha_{2}, \beta_{2}\right),
\end{array}
\end{equation}
where the conditional distributions $\left(a_{k} \middle\vert \tau_{a}^{-1}, \tau_{\epsilon}^{-1}, \mu,\{y_{jk}\}_{j},\{w_{jk}\}_{j},w_{k}\right)$ are independent and
$(\alpha_{1},\alpha_{2},\beta_{1},\beta_{2})$ are fixed hyperparameters.
Each line in Equation~\ref{gibbsscan} denotes a pseudo posterior distribution for one set of parameters conditioned on or fixing the others, which we refer to as a full conditional pseudo posterior distribution where $\left(c\middle\vert d\right)$ denotes a density for $c$ conditioned on $d$.

We utilize each full conditional distribution to demonstrate the frequentist consistency with respect to $\mathbb{P}_{\lambda_{0}},\mathbb{P}^{\pi}$. The proof for every parameter block, represented by a line Equation~\ref{gibbsscan}, follows four general steps: derive the full conditional pseudo posterior distribution of the parameter, which is analytically tractable, perform the derivation for the \emph{observed} sample, proceed to extract the first two moments of the full conditional pseudo posterior estimator, and expand them from the observed sample to the population by inserting the random inclusion indicators, $\delta$.  We, subsequently, compute expectations of these expanded summary measures of the full conditionals with respect to $\mathbb{P}_{\lambda_{0}},\mathbb{P}^{\pi}$ that we use to demonstrate the posterior contraction result in $L_{1}$-$\mathbb{P}_{\lambda_{0}},\mathbb{P}^{\pi}$ using the Chebyshev inequality.  In summary, the consistency proof approach for each parameter block in $\lambda$ demonstrates the contraction of each full conditional posterior distribution under $\mathbb{P}_{\lambda_{0}},\mathbb{P}^{\pi}$, which does not depend on the sampling mechanics to realize samples drawn from these distributions.

\subsection{Assumptions}\label{sec:assume}

Our theoretical setup is based on that of \citet{williams2020}. They demonstrate that consistent estimation of model parameters can be achieved by exponentiating the likelihood under mild conditions met by most sampling designs, such as balanced and unbalanced two-stage designs, that restrict the growth of the size of each cluster and usual restrictions on the population generating model; see simulation studies of Sections~\ref{sec:sims1} and \ref{sec:sims2}, respectively, for different examples of sampling designs. Their result also suggests that extending the exponentiation to include weights for latent variables which correspond to sampled units (clusters) in the population will also lead to consistent estimation. The simulation study results in Section \ref{sec:sims1} and the derivations sketched in \citet{SavitskyWilliams+2022+901+928} support this conclusion. However, the expanded simulations for extremely unbalanced designs in Section \ref{sec:sims2} reveal that additional conditions for within cluster sampling designs are needed to assert consistency. We use the one-way ANOVA as an opportunity to more directly examine these properties for a simple canonical case.

Some of our assumptions as well as our main result, stated in Theorem~\ref{thm1}, are asymptotic in nature; that is, we use phrases in the sequel such as ``for sufficiently large $M$", which is consistent with this approach.   Our asymptotic framing and use of such phrases is standard as evidenced by \citet{Ghosal00convergencerates}, the canonical work on the frequentist consistency of posterior distribution estimators.  We include rates of convergence in Theorem~\ref{thm1} in order to provide insight on the speed with which each result is achieved.

We enumerate assumptions on the sampling design distribution $\mathbb{P}^{\pi}$ and bounds on sampling quantities used in our proof statements of frequentist consistency of our Bayesian estimators derived from full conditional pseudo posterior distributions. Explicit assumptions on the population and estimation models, (A1)--(A3) in \cite{williams2020}, are not required here because we explicitly work with these formulations for the one-way ANOVA model.
\begin{description}
\item[\namedlabel{itm:sampling design}{(C4)}] (Sampling Design)
     \begin{description}
     \item[(i)\label{independentgroups}] $(w_{k},\mathbbm{1}\left\{k\in S_{c}\right\},S_{k} = s_{k})$ $(k = 1,\ldots,M)$  are mutually independent and
     may depend on $a_{k}$.
     \item[\namedlabel{withingroup}{(ii)}] $\mathbb{E}_{\mathbb{P}_{\lambda_{0}}}
     \left(\sum_{j \in S_{h}}w_{j|h}\epsilon_{j h0} / \sum_{j \in S_{h}}w_{j|h}\right)
     \rightarrow \mathbb{E}_{\mathbb{P}_{\lambda_{0} }}\left( \sum_{\ell = 1}^{N_h} \epsilon_{\ell h0} \right) / \mathbb{E}_{\mathbb{P}_{\lambda_{0} }}\left(\hat{N}_{h}\right)= 0$ for sampling designs where $\hat{N}_{h}=\sum_{j \in S_{h}}w_{j|h}$ is constant by construction \emph{or} for a within group sampling fraction, $f_{h} = n_{h}/N_{h} < 1$ sufficiently large.  In the first case, $\hat{N}_{h}$ is independent of $\mathbb{P}_{\lambda_{0}}$ and in the second it limits to independence.  We require that the stated expectation with respect to $\mathbb{P}_{\lambda_0}$ converges to $0$ for any $S_{h}\in \mathcal{S}_{h}$ and for every $ h \in (1,\ldots,M)$.
     \end{description}
\item[\namedlabel{itm:bounds}{(C5)}] (Bounds Governing Sampling Design)
    $m^{(0)}, n^{(0)}_{k},K_{0},K_{1},K_{2},K_{3},K_{4}$ are constants all $>0$, such that as $N\uparrow\infty$:
    \begin{description}
      \item[\namedlabel{constantfrac}{(o)}] $m = \mathcal{O}(M)$ such that $f_{c} = m/M = \mathcal{O}(1)$.
      \item[(i)] $ M \geq K_{4} N,~ M = \mathcal{O}(N)$.
      \item[\namedlabel{withinbounded}(ii)] $\forall k \in (1,\ldots,M), N_{k} \le K_{3}$
      \item[(iii)] $\forall k \in (1,\ldots,M),~1/K_{2} \leq w_{k}m^{(0)}/M \leq K_{2}$ with $P^{\pi}-$ probability $1$.
      \item[(iv)] $\forall (j,k) \in U,~1/K_{1} \leq w_{j\vert k}n_{k}^{(0)}/N_{k} \leq K_{1}$ with $P^{\pi}-$ probability $1$.
    \end{description}
\end{description}
Assumption~\ref{itm:sampling design}\ref{withingroup} arises in order to achieve an expectation for the average of weighted (posterior) estimated random effects, $M^{-1}_{k\in S_{c}} \sum w_{k}a_{k} = 0$, which in turn is required for consistency of $\tau_{a}^{-1}$ (Appendix \ref{sec:proof-prop-refpr}).  Assumption~\ref{itm:sampling design}\ref{withingroup} requires that weighted within cluster residuals for sampled units be balanced or sum to $0$ for all clusters.  If this condition is violated it leads to a biased result for the average of weighted random effects and, in turn, for the posterior distribution estimator of  $\tau_{a}^{-1}$ (and also of $\mu$ and $\tau_{\epsilon}$).

Without replacement, proportion-to-size (pps) sampling designs that utilize explicit stratification within cluster and assign higher within cluster inclusion probabilities, $\{w_{r | h}\}_{r}$ (where $r \in (1,\ldots,R)$ denotes stratum) to those strata with larger magnitude $\{y_{j|h}\}_{j \in r}$ would be expected to meet Assumption~\ref{itm:sampling design}\ref{withingroup}. For example, the Current Employment Statistics survey administered by the U.S. Bureau of Labor Statistics uses a stratified proportion-to-size sampling design. Similarly, pps designs that use sorting/implicit stratification would be expected to nearly meet this condition, meaning that the sum of residuals will be near to $0$ such that the one-way ANOVA estimator for $(\mu, \tau_{\epsilon}^{-1},\tau_{a}^{-1})$ may express a small magnitude bias.  The residual balance would converge on $0$ as the number of implied strata increases.

Focusing on the denominator of Assumption~\ref{itm:sampling design}\ref{withingroup}, it is generally true that $w_{j|h} \propto 1/\pi_{j|h}$ for units sampled within each cluster and we construct $\pi_{j|h}$ in Section 2 to be a function of $y_{j|h}$ that is, in turn, generated from $\mathbb{P}_{\lambda_{0}}$.  This means in the most general sense that $\hat{N}_{h}=\sum_{j \in S_{h}}w_{j|h}$ would be expected to depend on $\mathbb{P}_{\lambda_{0}}$ such that Assumption~\ref{itm:sampling design}\ref{withingroup} would be violated (though the summation over within cluster sampling weights would typically attenuate the dependence).  Yet, $\hat{N}_{h}$ would be constant (and thus independent of $\mathbb{P}_{\lambda_{0}}$), for designs that use stratification.  In the case of simple random sampling within cluster, the inclusion probabilities are constant such that $\hat{N}_{h}$ is constant by design.  It is easy to verify that the same is true for stratified random sampling with simple random sampling (and constant inclusion probabilities) within each stratum. Similarly, proportion-to-size (pps) sampling designs that employ strata, assigning larger inclusion probabilities to higher magnitude strata, would also produce a constant $\hat{N}_{h}$. Finally, pps designs that use implicit stratification or (systematic) sorting would produce a nearly constant value for $\hat{N}_{h}$.

Focusing on the required residual balance in the numerator of Assumption~\ref{itm:sampling design}\ref{withingroup}, as the number of strata used within each cluster increases to $5-10$, any sample drawn from each cluster would contain both positive and negative residuals.   One does not need an equal balance of residuals between positive and negative for sampled units within each cluster because the weighting of each each within cluster residual, $\epsilon_{jh0}$, by $w_{j|h}$ will push the weighted sum to balance in residuals such that the sum is $0$ or nearly so.

If the within cluster sampling fraction, $f_{h}$, becomes large (in the range of $0.5$) a pps design that does \emph{not} use either of implicit or explicit stratification would still nearly obey Assumption~\ref{itm:sampling design}\ref{withingroup} because the within cluster weighted sum of residuals would be nearly balanced at $0$ since the larger sample size makes more likely that the sampled values in every cluster would contain both positive and negative residuals.  Under a larger-valued $f_{h}$, $\hat{N}_{h}$ would be nearly constant because the variance of the sampling weights declines (which stabilizes their sum) until each $w_{j|h}$ converges to a constant in the limit as $f_{h}$ limits to $1$.

We construct simulation studies in Sections~\ref{sec:sims1} and ~\ref{sec:sims2} that explicitly evaluate the consistency for estimation of $(\mu, \tau_{\epsilon}^{-1},\tau_{a}^{-1})$ for a population generated from the one-way ANOVA distribution under classes of sampling designs that are known to both meet and violate Assumption~\ref{itm:sampling design}\ref{withingroup}. We show that when this key assumption is met that our sampling weighted estimator is consistent and when the assumption is violated that consistency is not achieved.  We further study sampling designs that slightly violate  Assumption~\ref{itm:sampling design}\ref{withingroup} (e.g., by having a small, non-zero within cluster weighted sum of residuals) to study the sensitivity of the resulting estimator consistency.

Assumption~\ref{itm:bounds}\ref{constantfrac} requires the cluster sampling fraction limit to a constant strictly greater than $0$.  Such is required in order for the weighted sum of sampled random effects $M^{-1} \sum_{k\in S_{c}} w_{k}a_{k} = 0$, which in turn is needed for consistency of the estimated random effects variance, $\tau_{a}^{-1}$. 

By contrast, \citet{slud_2019} shows in Lemma 11 that the MLE for Equations~\ref{eq:sampobsmodel}  is \emph{not} consistent if $f_{c}$ limits to $0$ \emph{and} there is at least one cluster sampling weight, $w_{h},~h\in(1,\ldots,M)$ that limits to $\infty$.  In particular, they require $\frac{1}{N\tau_{\epsilon}}\times \mathop{\sum}_{k=1}^{m}\left[\frac{w_{k}N_{k}\tau_{\epsilon}^{-1}}{\tau_{\epsilon}^{-1} + N_{k}\tau_{a}^{-1}} - 1\right]$ be bounded away from $0$ (for any realized sample of clusters indexed by $k\in(1,\ldots,m)$) for Equation~\ref{eq:sampobsmodel} to be \emph{inconsistent}.  Since there is a strong $1/N$ pull to $0$ in that term, \citet{slud_2019} require that $w_{k}$ limit to $\infty$ at a rate faster than $\mathcal{O}(M)$ to counteract this tendency towards consistency.  Their inconsistent result, then, is something of a ``corner case" in that $\pi_{h}$ is bounded away from $0$ (which results in $w_{h}$ be finite (bounded from above by a constant)) for most practical sampling designs, even when the cluster sampling fraction is small \citep{2015arXiv150707050S}. 

We construct a simulation study in Appendix~\ref{sec:sims3} that, indeed, shows our Bayesian estimator is \emph{inconsistent} if the cluster sampling fraction, $f_{c}$ limits to $0$ \emph{and} the $w_{h}$ limit to $\infty$ at a non-linear $\mathcal{O}(\frac{M^2}{m})$ for some $h$. By contrast, we discovered that even if the cluster sampling fraction limits to $0$ our simulation experiments show that there is a very small-to-negligible estimation bias in practice if the $w_{h}$ grow at a linear $\mathcal{O}(\frac{M}{m})$ rate under sampling designs that obey Assumption~\ref{itm:sampling design}\ref{withingroup}. 

Stepping back, it is our experience in practice that $f_{c}$ converges to a constant bounded away from $0$ (please see \citet{2015arXiv150707050S} in the case of a single-level model) and that the use of an assumption that it converges to $0$ is done for convenience (e.g., to simplify variance estimation).

Using \ref{itm:bounds} and Lemma 5 of \citet{slud_2019}, the weights are bounded: $w_{k} \le K_{2} \frac{M}{m^{(0)}} < \gamma_{1}$,
$w_{j|k} \le K_{1} \frac{N_{k}}{n_{k}^{(0)}} < \gamma_{2}$, where $0 <(\gamma_{1},\gamma_{2}) < \infty$ and therefore $w_{jk} \le \gamma_{1} \gamma_{2} \equiv \gamma < \infty$.
We note that there is no additional restriction on informativeness of the sampling design at any stage. However, condition \ref{itm:sampling design}\ref{withingroup} will tend to dampen extremely skewed designs by requiring a larger sampling fraction. We explicitly bound the cluster size $N_{k}$.

We note that these conditions are also very similar to the conditions needed for consistency of single-level models \citep[For example (A4) and (A5) in][]{williams2020} with the addition of condition \ref{itm:sampling design}\ref{withingroup} for the two-level models.


\subsection{Weighted Sum of Random Effects}\label{sec:sumre}

We begin with a derivation of the full conditional pseudo posterior distribution for $a_{k}$. Noting that $\{a_{k}\}$ are \emph{a posteriori} conditionally independent,
\begin{equation}\label{ak-post-1}
\begin{array}{rl}
  \left(a_{k}\middle\vert \{y_{jk}\}, \{w_{jk}\} , \{w_{k}\}, \tau_{\epsilon}^{-1}, \tau_{a}^{-1}\right) \propto \exp\left(a_{k}\left\{\tau_{\epsilon}\mathop{\sum}_{j\in S_{k}}w_{jk}\tilde{y}_{jk}\right\} - \frac{1}{2}a_{k}^{2} \left\{ \tau_{\epsilon}\mathop{\sum}_{j\in S_{k}}w_{jk}\ + \tau_{a}w_{k}\right\}\right),
\end{array}
\end{equation}
where $\tilde{y}_{jk} = y_{jk} - \mu$.  In Section \ref{sec:mu}, we demonstrate that $\mu$ contracts on $\mu_{0}$ such that we replace $\mu$ with $\mu_{0}$ for $M$ sufficiently large. Then $\tilde{y}_{jk} \rightarrow y_{jk} - \mu_0 = a_{k0} + \epsilon_{jk0}$. From this algebraic reduction, we obtain that
\begin{equation}\label{ak-post-2}
\left(a_{k}\middle\vert\{y_{jk}\}, \{w_{jk}\} , \{w_{k}\}, \tau_{\epsilon}^{-1}, \tau_{a}^{-1}\right) = \mathcal{N}\left(a_{k}\middle\vert h_{k},\phi_{k}^{-1}\right),
\end{equation}
where $e_{k} = \tau_{\epsilon}\mathop{\sum}_{j\in S_{k}}w_{jk} (a_{k0} + \epsilon_{jk0})$, $\phi_{k} = \tau_{\epsilon}\mathop{\sum}_{j\in S_{k}}w_{jk}\ + \tau_{a}w_{k}$ and $h_{k} = \phi_{k}^{-1}e_{k}$.

The random effects $\{a_{k}\}$ are unobserved, but the full conditional of $\tau^{-1}_{a}$ in Equation \ref{gibbsscan} depends on them through a weighted average of $\{a_{k}\}$; therefore, we proceed to formulate our estimator of the weighted average as a conditional distribution of the weighted average of random effects as
\begin{equation}\label{eq:wtd-avg}
  \left(M^{-1}\sum_{k \in S_{C}} w_k a_{k} \middle\vert \{y_{jk}\}, \{w_{jk}\} , \{w_{k}\}, \tau_{\epsilon}^{-1}, \tau_{a}^{-1}\right) = \mathcal{N}\left(M^{-1}\sum_{k \in S_{C}} w_k a_{k} \middle\vert M^{-1}\sum_{k \in S_{C}} w_k h_{k}, M^{-2}\sum_{k \in S_{C}} w_k^{2} \phi_{k}^{-1}\right),
\end{equation}
where the mean and variance of the total are simple weighted sums of the mean and variance of the individual conditional distributions for $\{a_k\}$.  We focus on a weighted average estimator because consistency of this estimator to the (unweighted) average of true random effects parameters for the population is all that we need to demonstrate the consistency of the generating random effects variance, $\tau_{a}^{-1}$, that we perform in the sequel. The parameter $\tau_{a}^{-1}$ only depends on the aggregated collection of the $\{a_k\}$, since it is a generating hyperparameter under Equation~\ref{poptruth}.

Assume that the population of random effects is generated under the model
\begin{equation}
  a_{10},\ldots,a_{M0} \iid \mathbb{P}_{\tau_{a0}} \propto \mathbb{P}_{\theta_{0}},
\end{equation}
for $\tau_{a0}\in \theta_{0} \in \Theta \subset \mathbb{R}^{3}$ with density, $p_{\theta} d \theta = p(a_h|\theta)d\theta$. The random effects, $\mathbf{a}_{0} \in A$ and $\mathbb{P}_{\theta_{0}}$ are defined on the measurable space, $(A,\mathcal{M}(A))$, where $\mathcal{M}(A)$ denotes the $\sigma$-algebra of measurable sets of $A$.
The random effects are \emph{latent} and \emph{estimated as parameters} in the first/top level of a hierarchical Bayesian estimation model stated in Equation~\ref{popmodel} using the prior distribution $\mathbb{G}_{\theta}$ resulting in the posterior density
\begin{equation}
  g\left(a_{h}\mid\theta,\{y_{\ell h}\}_{\ell}\right) \propto \left[\mathop{\prod}_{\ell=1}^{N_{h}}p(y_{\ell h}|\theta,a_{h}) \right] \times g\left(a_{h}|\tau_{a}^{-1}\right),
\end{equation}
where $\mathbb{G}^{y}_{\theta}$ is the corresponding posterior distribution governing $\mathbf{a} \in A$ such that $\mathbb{G}_{\theta}$ and $\mathbb{G}^{y}_{\theta}$ are also defined on $(A,\mathcal{M}(A))$ under the generating model for $(Y_{1 1},\ldots,Y_{N_{M} M}) \ind \mathbb{P}_{\lambda_{0}}$ of Equation~\ref{poptruth}. The following proposition is crucial to our theory development later.
\begin{proposition}\label{prop1}
  Let $q(\mathbf{a}_{0}) = M^{-1}\sum_{h=1}^{M}a_{h0}$ be a summary function defined on the domain of the population random effects. Then, the statistic based on the observed random effects
  $q^{\pi}(\mathbf{a}) = M^{-1}\sum_{h=1}^{M}\delta_{h}w_{h}a_{h}$ converges to $q(\mathbf{a}_{0})$ in $L_1$-norm with respect to the $\mathbb{P}_{\lambda_{0}},\mathbb{P}^{\pi}$ distribution as $M \rightarrow \infty$.
\end{proposition}
The proof of this proposition is in the appendix with other proofs.

An important step in the proof of Proposition \ref{prop1} is to identify the distribution governing the randomness in $q^{\pi}(\mathbf{a})$ is the joint distribution $\mathbb{G}_{\theta_0}^{y}, \mathbb{P}^{\pi}$.  We use this idea in Theorem \ref{thm1} presented later to show that the variance parameter, $\tau_{a}^{-1}$, in the prior distribution for $a_{1},\ldots,a_{M}$ contracts in probability under the distribution governing the \emph{second} level of the Bayesian hierarchical probability estimation model, $\mathbb{G}^{y}_{\theta_{0}},\mathbb{P}^{\pi}$, to $\tau_{a0}$; for example, let $r^{\pi} = \mathbb{E}(\tau_{a}^{-1}|(\delta_{h}),(w_{h}),\mathbf{a})$, the pseudo posterior mean for $\tau_{a}^{-1}$ under the formulation of Equation~\ref{eq:sampcompletemodel}. We show,
\begin{equation}
\begin{array}{rl}
\mathbb{E}_{\mathbb{G}^{y}_{\theta_{0}},\mathbb{P}^{\pi}}\left[r^{\pi}(\mathbf{a})\right]  &= \mathbb{E}_{\mathbb{G}^{y}_{\theta_{0}}}\left[\mathbb{E}_{\mathbb{P}^{\pi}}\left(r^{\pi}(\mathbf{a})|\theta_{0}\right)\right]\\
&= \ \mathbb{E}_{\mathbb{G}^{y}_{\theta_{0}}}\left[L(q(\mathbf{a}))\right]\\
&\stackrel{L_{1}-\mathbb{P}_{\lambda_{0}},\mathbb{P}^{\pi}}{\longrightarrow} \mathbb{E}_{\mathbb{P}_{\theta_{0}}}\left[L(q(\mathbf{a}_{0}))\right],
\end{array}
\end{equation}
for any continuous function $L(\cdot)$.

Proposition \ref{prop1} has two important implications. Firstly, our results presented in the sequel only require consistency of the \emph{average} of the random effects, which is proved in Proposition \ref{prop1}, to establish consistency of the random effects generating variance, $\tau_{a}^{-1}$; in particular, we do not require consistency of \emph{each} individual random effect, $a_{h}$. Secondly, the random effects, $\mathbf{a}$, in the first level of the Bayesian hierarchical estimation model are input into a summary function whose output may be used for estimating  $\tau_{a}^{-1}$ in the second level of the hierarchical model; however, $\mathbf{a}$ are estimates of $\mathbf{a}_{0}$, both of which are unobserved. We use Proposition \ref{prop1} to show the  concentration of the posterior distribution of the summary function defined using $\mathbf{a}$ under the population generating distribution, $\mathbb{P}_{\lambda_{0}},\mathbb{P}^{\pi}$ (a distribution that is random with respect to $\{y_{jk}\}_{jk}$).

\subsection{Main Results for $\theta$}\label{sec:mu}

We first derive the full conditional pseudo posterior distributions of the intercept and precision parameters for the random cluster effects and idiosyncratic noise, respectively. Equation \ref{obssampmodel} implies that the three conditional pseudo posterior distributions have densities
\begin{equation} \label{prec-tau}
  \begin{array}{rl}
    \left( \mu\middle\vert\{y_{jk}\}, \{a_{k}\}, \{w_{jk}\}, \tau_{\epsilon} \right) \ & \propto \left[
\prod_{k=1}^{m} \prod_{j=1}^{n_{k}} \mathcal{N}\left(\tilde{y}_{jk}\middle\vert\mu, \tau_{\epsilon}\right)^{w_{jk}}
\right] \times \mathbbm{1}\{\mu \in (-\infty, \infty)\}\\
\ & = \ \mathcal{N}\left (\dfrac{\sum_{k = 1}^{m} \sum_{j = 1}^{n_{k}} w_{jk}\tilde{y}_{jk}}{\sum_{k = 1}^{m} \sum_{j = 1}^{n_{k}} w_{jk}}, \left[\tau_{\epsilon}\sum_{k = 1}^{m} \sum_{j = 1}^{n_{k}} w_{jk}\right]^{-1} \right) \\
\left (\tau_{a}^{-1}\middle\vert\{a_{k}\}, \{w_{k}\}  \right)   & \propto \ \left[\prod_{k=1}^{m}
	\mathcal{N}(a_{k} \vert 0, \tau_{a}^{-1})^{w_{k}}
\right] \mathcal{IG}(\tau_{a}^{-1} \vert \alpha_1,\beta_1) \\
& = \ \mathcal{IG}(\frac{1}{2} \sum_{k = 1}^{m}  w_{k} + \alpha_1, \frac{1}{2} \sum_{k = 1}^{m} w_{k} a_{k}^{2} + \beta_1),\\
\left (\tau_{\epsilon}^{-1}\middle\vert\{y_{jk}\}, \{a_{k}\}, \{w_{jk}\} , \mu \right)   & \propto \ \left[\prod_{k = 1}^{m} \prod_{j = 1}^{n_{k}}
	\mathcal{N}(\epsilon_{jk0}|0, \tau_{\epsilon}^{-1})^{w_{jk}}
\right] \mathcal{IG}(\tau_{\epsilon}^{-1}|\alpha_2,\beta_2) \\
& = \ \mathcal{IG}\left(\dfrac{1}{2} \sum_{k = 1}^{m} \sum_{j = 1}^{n_{k}} w_{jk} + \alpha_{2}, \dfrac{1}{2} \sum_{k = 1}^{m} \sum_{j = 1}^{n_{k}} w_{jk}\epsilon_{jk0}^2 + \beta_2\right),
\end{array}
\end{equation}
where $\tilde{y}_{jk} = y_{ij} - a_{k}$ and $\epsilon_{jk0} = y_{jk} - \mu_{0} - a_{k0}$.

We use means and variances of the conditional posterior distributions in Equation \ref{prec-tau} as summary measures for deriving their posterior concentration. The conditional
conditional pseudo posterior $\left( \mu\middle\vert\{y_{jk}\}, \{a_{k}\}, \{w_{jk}\}, \tau_{\epsilon} \right)$ is Gaussian with its mean and variance specified in Equation \ref{prec-tau}.
The expected values of the conditional pseudo posteriors $\left (\tau_{a}^{-1}\middle\vert\{a_{k}\}, \{w_{k}\}  \right)$ and $\left (\tau_{\epsilon}^{-1}\middle\vert\{y_{jk}\}, \{a_{k}\}, \{w_{jk}\} , \mu \right)$ are
\begin{align}
  \label{eq:1}
  \mathbb{E} \left (\tau_{a}^{-1}\middle\vert\{a_{k}\}, \{w_{k}\}  \right) &=
                                                                             \dfrac{\sum_{k = 1}^{m} w_{k} a_{k}^{2} + 2\beta_1}{\sum_{k = 1}^{m}  w_{k} + 2\alpha_1-2}
                                                                             \approx \dfrac{\sum_{k = 1}^{m} w_{k} a_{k}^{2}}{\sum_{k = 1}^{m}  w_{k}}
                                                                             =\frac{1}{\hat{M}}  \sum_{k = 1}^{m} w_{k} a_{k}^{2}, \nonumber\\
  \mathbb{E}\left (\tau_{\epsilon}^{-1}\middle\vert\{y_{jk}\}, \{a_{k}\}, \{w_{jk}\} , \mu \right) &=
                                                                                                     \dfrac{\sum_{k = 1}^{m} \sum_{j = 1}^{n_{k}} w_{jk}\epsilon_{jk0}^2  + 2 \beta_2}{ \sum_{k = 1}^{m} \sum_{j = 1}^{n_{k}} w_{jk}  + 2\alpha_2-2} 
 \approx 
   \frac{1}{\hat{N}}  \sum_{k = 1}^{m} \sum_{j = 1}^{n_{k}} w_{jk}\epsilon_{jk0}^2.
\end{align}
where the constants $\beta_1, \beta_2$ and $\alpha_1, \alpha_2$ are negligible, because $m = \mathcal{O}(M) = \mathcal{O}(N)$. The variance of the conditional pseudo posteriors $\left (\tau_{a}^{-1}\middle\vert\{a_{k}\}, \{w_{k}\}  \right)$ and $\left (\tau_{\epsilon}^{-1}\middle\vert\{y_{jk}\}, \{a_{k}\}, \{w_{jk}\} , \mu \right)$ satisfy
\begin{align}
  \label{eq:2}
  \mathbb{V} \left (\tau_{a}^{-1}\middle\vert\{a_{k}\}, \{w_{k}\}  \right)
&\le \dfrac{C_{1}\left[\sum_{k = 1}^{m} w_{k} a_{k}^{2}\right]^2}{\left[\sum_{k = 1}^{m}  w_{k}\right]^3}
  =\dfrac{C_{1}\left[ \sum_{k = 1}^{m} w_{k} a_{k}^{2} \right]^{2}}{\hat{M}^3}, \\
\mathbb{V}\left (\tau_{\epsilon}^{-1}\middle \vert \{y_{jk}\}, \{a_{k}\}, \{w_{jk}\} , \mu \right)
&\le \dfrac{C_{2}\left[\sum_{k = 1}^{m} \sum_{j = 1}^{n_{k}} w_{jk}\epsilon_{jk0}^2\right]^2}{\left[\sum_{k = 1}^{m} \sum_{j = 1}^{n_{k}} w_{jk}\right]^3}
=\dfrac{C_{2}\left[ \sum_{k = 1}^{m} \sum_{j = 1}^{n_{k}} w_{jk}\epsilon_{jk0}^2 \right]^{2}}{\hat{N}^3},  \nonumber
\end{align}
For simplicity, we can invoke Slutsky's theorem to proceed with $M$ and $N$ instead of $\hat{M}$ and $\hat{N}$, respectively.

Our main result is about the concentration of the conditional posterior distributions in Equation \ref{prec-tau}. Studying their asymptotic behavior is nontrivial because the random effects $\{a_h\}$ are unknown. Following the discussion after Proposition \ref{prop1}, we address this issue by replacing $\sum_{h=1}^{M}\delta_{h}w_{h}a_{h}$ with $\sum_{h=1}^{M}a_{h0}$ in Equations \ref{eq:1} and \ref{eq:2} for a sufficiently large $M$. With some additional assumptions and moment conditions on the population model, we have the following theorem.
\begin{theorem}\label{thm1}
  If Assumptions \ref{itm:sampling design} and \ref{itm:bounds} hold and $M$ and $N$ are sufficiently large, then
  \begin{enumerate}
  \item $\mathbb{E}_{\mathbb{P}_{\lambda_{0}}, \mathbb{P}^{\pi}}\left[ \mathbb{E} (\mu\middle\vert\{y_{\ell h}:(\ell,h)\in S\}, \{a_{h}:h\in S_{c}\}, \{\delta_{\ell h}\}, \{w_{\ell h}:(\ell,h)\in S\}, \tau_{\epsilon}   \right] = \mu_{0}$;
  \item $\mathbb{E}_{\mathbb{P}_{\lambda_{0}}, \mathbb{P}^{\pi}}\left[\mathbb{V} (\mu\middle\vert \{y_{\ell h}:(\ell,h)\in S\}, \{a_{h}:h\in S_{c}\}, \{\delta_{\ell h}\}, \{w_{\ell h}:(\ell,h)\in S\}, \tau_{\epsilon}  \right] = \mathcal{O} (N^{-1})$;
  \item $\mathbb{E}_{\mathbb{G}^{y}_{\theta_{0}}, \mathbb{P}^{\pi}}\left[ \mathbb{E} \left (\tau_{a}^{-1}\middle\vert\{a_{h}:h\in S_{c}\}, \{\delta_{h}\}, \{w_{h}:h\in S_{c}\}   \right) \right] = \tau_{a0}^{-1}$;
  \item $\mathbb{E}_{\mathbb{P}_{\lambda_{0}}, \mathbb{P}^{\pi}}\left[ \mathbb{E} \left (\tau_{\epsilon}^{-1}\middle\vert \{y_{jk}\}, \{a_{k}\}, \{w_{jk}\} , \mu \right) \right] = \tau_{\epsilon 0}^{-1}$;
  \item $\mathbb{E}_{\mathbb{G}^{y}_{\theta_{0}}, \mathbb{P}^{\pi}}\left[ \mathbb{V} \left (\tau_{a}^{-1}\middle\vert\{a_{h}:h\in S_{c}\}, \{\delta_{h}\}, \{w_{h}:h\in S_{c}\}   \right) \right] = \mathcal{O}(M^{-1})$; and
  \item $\mathbb{E}_{\mathbb{P}_{\lambda_{0}}, \mathbb{P}^{\pi}}\left[ \mathbb{V} \left (\tau_{\epsilon}^{-1}\middle\vert \{y_{jk}\}, \{a_{k}\}, \{w_{jk}\} , \mu \right) \right] = \mathcal{O}(N^{-1})$.
  \end{enumerate}
\end{theorem}
The proof of this theorem is in the appendix. The Chebyshev's inequality and this theorem together imply the conditional pseudo posteriors $\left(\mu\middle\vert \{a_{k}\}, \{w_{k}\}, \tau_{\epsilon}\right)$, $\left (\tau_{a}^{-1}\middle\vert\{a_{k}\}, \{w_{k}\}  \right)$ and $\left (\tau_{\epsilon}^{-1}\middle\vert\{y_{jk}\}, \{a_{k}\}, \{w_{jk}\} , \mu \right)$ contract on $\mu_{0}, \tau_{a 0}^{-1}$ and $\tau_{\epsilon 0}^{-1}$, respectively, in $L_{1}$-$\mathbb{P}_{\lambda_{0}}, \mathbb{P}^{\pi}$ as $M, N $ tend to $\infty$.  For example, applying Theorem~\ref{thm1} and Chebyshev for $\mu$ produces,
\begin{equation*}
  \mathbb{E}_{\mathbb{P}_{\lambda_{0}}, \mathbb{P}^{\pi}}
\left[
\mathbb{P} \left( \abs{\mu - \mu_{0}}  > \delta \mid \cdot\right)\right]
 \le 
\dfrac{\mathbb{E}_{\mathbb{P}_{\lambda_{0}}, \mathbb{P}^{\pi}}
\left[\mathbb{V}  \left( \mu\middle\vert\cdot\right) \right]}{\delta^2}  = \mathcal{O}(N^{-1}).
\end{equation*}
In particular, the outer expectation, $\mathbb{E}_{\mathbb{P}_{\lambda_{0}}, \mathbb{P}^{\pi}}$ is random with respect to the data, $\{y_{jk}\}_{jk}$, while $\mathbb{P}(\mu\mid\cdot)$ and $\mathbb{V}(\mu\mid\cdot)$ are shorthand notations for the full conditional posterior distribution and variance, respectively, where the conditioning on parameters, sampling weights and data is implied.  These latter quantities are random with respect to parameter, $\mu$, such that the outer expectation provides the $L_{1}$-$\mathbb{P}_{\lambda_{0}}, \mathbb{P}^{\pi}$ frequentist convergence of the full conditional posterior probabibility for $\mu$ to $\mu_{0}$.  Please see Appendix~\ref{sec:proof-prop-refpr} for a more detailed exposition of the use of Chebyshev.  Theorem~\ref{thm1} tells us that the rate of convergence to $\mu_{0}$ is $\mathcal{O}(N^{-1})$.

Some comments are in order about the computation of two moments for all the three full conditionals in Theorem \ref{thm1}. First, the parameters of the full conditional  pseudo posterior distributions are random, and we assess their $L_1$ contraction with respect  to the joint distribution of population data generation and the taking of a sample, jointly, where the data and sample inclusion indicators are treated as random.  Second, the full conditional of $\tau_a^{-1}$ depends on the sample inclusion indicators $(\delta_h)$ and random effects $\{a_h\}$. This statistic is defined in Proposition  \ref{prop1}, so the $L_1$ contractions of this conditional is computed with respect to $\mathbb{G}^y_{\theta_0}, \mathbb{P}^{\pi}$ distribution jointly, where $\mathbb{G}^y_{\theta_0}$ is the population  distribution of $\{a_h\}$. Third, the population distributions of $\{y_{\ell h}\}_{(\ell h)}$ and $\{\delta_{\ell, h}\}_{(\ell h)}$ are $\mathbb{P}_{\lambda_0}$ and $\mathbb{P}^{\pi}$, respectively.  The  full conditionals of $\tau_{\epsilon}^{-1}$ and $\mu$ depend on $\{y_{\ell h}\}_{(\ell h)}$ and $\{\delta_{\ell, h}\}_{(\ell h)}$, so their $L_1$ contraction is computed with respect to their population distribution $\mathbb{P}_{\lambda_0}, \mathbb{P}^{\pi}$ jointly.  Finally, Theorem \ref{thm1} establishes joint model and design consistency through using our population distribution $\mathbb{P}_{\lambda_0}, \mathbb{P}^{\pi}$.

\subsection{Integrated Likelihood}\label{sec:ilike}

We may marginalize over $a_k$ for each $k \in (1,\ldots,m)$ from the augmented pseudo likelihood for $(\mu,\tau_{a}^{-1},\tau_{\epsilon}^{-1})$ with,

\begin{equation}\label{eq:ilike}
\begin{array}{rl}
     & L\left[(\mu,\tau_{a},\tau_{\epsilon})\middle\vert\{y_{jk}\}_{j\in S_{k}}\right]  \\
     & = \displaystyle\int_{a_{k} \in \mathbb{R}}\left[\prod_{j\in S_{k}}\mathcal{N}\left(\tilde{y}_{jk}\middle\vert a_{k},
     \tau_{\epsilon}^{-1}\right)^{w_{jk}}\right] \times \mathcal{N}\left(a_{k}\middle\vert 0,\tau_{a}^{-1}\right)^{w_{k}}d a_{k} \\
     & = \dfrac{\sqrt{2\pi}}{\phi_{k}^{\frac{1}{2}}}\exp\left[\frac{1}{2}\phi_{k}h_{k}^{2}\right] \times \dfrac{\tau_{\epsilon}^{\frac{1}{2}\sum_{j\in S_{k}} w_{jk}}\tau_{a}^{\frac{1}{2} w_{k}}}{(2\pi)^{\frac{1}{2} (n_{k}+1)}}\exp\left[-\frac{1}{2}\tau_{\epsilon}\sum_{j\in S_{k}} w_{jk}\tilde{y}_{jk}^{2}\right]\\
     & = \dfrac{1}{\mathcal{N}\left[h_{k}(\tau_{\epsilon},\tau_{a},\mu)\middle\vert 0,\phi_{k}(\tau_{\epsilon},\tau_{a})^{-1}\right]} \times \dfrac{\tau_{a}^{\frac{1}{2} w_{k}}}{\sqrt{2\pi}} \times \dfrac{\tau_{\epsilon}^{\frac{1}{2}\sum_{j\in s_{k}}w_{jk}}}{(2\pi)^{\frac{1}{2} n_{k}}}\exp\left[-\frac{1}{2}\tau_{\epsilon}\sum_{j\in S_{k}}w_{jk}\left(y_{jk}-\mu\right)^{2}\right],
\end{array}
\end{equation}
where $\phi_{k}(\tau_{\epsilon},\tau_{a}) = \tau_{\epsilon}\sum_{j\in S_{k}}w_{jk} + \tau_{a} w_{k}$, $~e_{k}(\tau_{\epsilon},\mu) = \tau_{\epsilon}\sum_{j\in S_{k}}w_{jk}(y_{jk}-\mu)$, and $h_{k}(\tau_{\epsilon},\tau_{a},\mu) = e_{k}(\tau_{\epsilon},\mu) / \phi_{k}(\tau_{\epsilon},\tau_{a})$. The form of the integrated likelihood is the quotient of a proper normal distribution multiplied by two sampling-weighted improper normal distribution kernels.  This integrated pseudo likelihood is a valid (integrable) distribution function under an improper prior for $\mu \propto 1$ because it is straightforward to show that the last improper normal distribution may be normalized to a proper normal distribution with precision, $\tau_{\epsilon}\sum_{j\in S_{k}}w_{jk}$.

After multiplying all $m$ integrated likelihoods, Equation \ref{eq:ilike} may be used to sample $(\mu, \tau_{a}^{-1},\tau_{\epsilon}^{-1})$, all under prior distributions earlier specified.    None of the full conditional  distributions admit closed-form distributions that we may use to assess consistency of the estimators.  
To demonstrate consistency, we can appeal to the equivalence between the augmented approach, with consistency arguments presented above. Further, we utilize simulations like those in Sections \ref{sec:sims1} and \ref{sec:sims2} to provide insight into the performance of the integrated pseudo posterior distributions under Bayesian estimation, in which the posterior estimates from both the augmented and integrated likelihood are indistinguishable to several digits of precision.

\section{Simulation Study 1: Strongly Informative Balanced Design for $\sigma_\epsilon$ and $\sigma_a$}\label{sec:sims1}
In this section, we demonstrate the estimation properties of several alternative methods applied to an extremely informative design in which both the random effects and the random noise have very different distributions between the population and each realized sample. In particular, we compare the double-weighting approach to single-weighting and to equal weighting each using the augmented data approach of Equation \ref{eq:sampcompletemodel}. We also compare to the EM algorithm of \citet{slud_2019} and the pairwise composite method of \citet{yi:2016}. We further demonstrate that for the simple canonical one-way ANOVA, several options for implementing estimation of the double-weighting scheme are available and produce similar results: MCMC sampling of the augmented model (\ref{eq:sampcompletemodel}), MCMC sampling using the integrated likelihood (\ref{eq:sampobsmodel}), and maximizing the posterior using the integrated likelihood.

\subsection{Model}\label{sec:sims1model}
For the population, we generate values $y_{h \ell} = b_0 + a_{h} + \epsilon_{h \ell}$ such that $a_{h} \sim N(0,\sigma_a)$ and
$\epsilon_{h \ell} \sim N(0, \sigma_{\epsilon})$, for clusters $h = 1, \ldots, M$ and individuals $\ell = 1, \ldots, N_{h}$.

\subsection{Sampling and Estimation}\label{sec:sims1samp}
We use population parameters $\{ b_0 = 1, \sigma_a = 2, \sigma_{\epsilon} = 3\}$. For each of three population sizes, we do the following:
\begin{enumerate}
\item Generate R = 100 populations from the one-way ANOVA distribution (using true values, $\mu_{0} = 1$, $\tau_{a}^{-1} = 2$, $\tau_{\epsilon}^{-1} = 3$).
	\begin{enumerate}
		\item With $M = \{1000,2000,4000\}$ clusters
		\item Each with $N_{h} = \{40 , 40 , 40 \}$ individuals in each cluster
	\end{enumerate}
\item For each $r = 1,\ldots, R$ population, draw a two-stage sample via `mstage'  in R \citep{sampling}
	\begin{enumerate}
		\item Sample $m = \{50,200,800\}$ clusters.
		\begin{enumerate}
			\item Using systematic PPS sampling with size $\pi_h	\propto (a^2_{h} + 1)$ 	
		\end{enumerate}
		\item Sample $n_{k} = \{5,5,5\}$ individuals in each cluster
		\begin{enumerate}
			\item Using systematic PPS sampling with size $\pi_{\ell|h}	\propto (\epsilon^2_{h \ell} + 1)$ 	
		\end{enumerate}
	\end{enumerate}
\item For each $r = 1,\ldots, R$ sample, estimate $\{b_0, \sigma_a, \sigma_{\epsilon}\}$
	\begin{enumerate}
		\item Using equal weights $w_{k} = 1$,  $w_{k j} = 1$ via `lmer' in R \citep{lme4}
		\item Using single weights $w_{k} = 1$,  $w_{k j} \propto 1/\pi_{k j}$ via Stan \citep{stan:2015}
		\item Using double weights $w_{k} \propto 1/\pi_{k}$,  $w_{k j} \propto 1/\pi_{k j}$ via Stan \citep{stan:2015}.
		\item Using double weights $w_{k} \propto 1/\pi_{k}$,  $w_{k j} \propto 1/\pi_{k j}$ in an EM algorithm as in \citet{slud_2019}
		\item Using two stage weights $w_{k} \propto 1/\pi_{k}$,  $w_{j|k} \propto 1/\pi_{j|k}$ and the pairwise composite method of \cite {yi:2016} as implemented in `svylme' in R \citep{svylme}.
	\end{enumerate}
\end{enumerate}

The EM algorithm in \citet{slud_2019} assumes non-informative within cluster sampling to derive the iterative formula for the EM algorithm.  By contrast, the data augmented Bayes formulation does not require closed form expressions for full conditionals (though they exist in the simple case of the one-way ANOVA used here).   Even more, the Bayes approach does not suppose any specific posterior sampling algorithm.
In this simple case, it is possible to integrate out the random effects and perform Bayesian inference on the integrated (or observed) likelihood. While the more general approach is data augmentation, we also compare to this equivalent alternative. Section \ref{sec:consistency} contains details on the convergence using the data augmented approach and includes the specification of the integrated likelihood (Section \ref{sec:ilike}).

\subsection{Results}\label{sec:results}
\subsubsection{Sample Properties}

\begin{figure}
\centering
\includegraphics[width = 0.95\textwidth,
		page = 1,clip = true, trim = 0.0in 0.0in 0in 0.in]{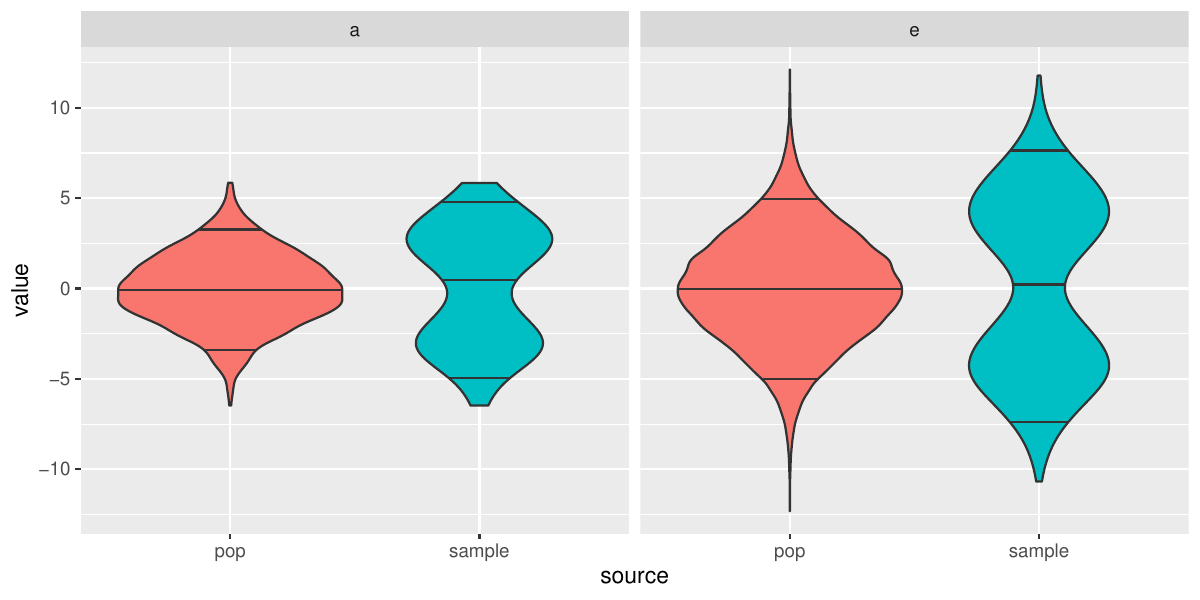}
\caption{Distributions and quantiles (5\%, 50\%, 95\%) of random effects $\mathbf{a}$ and noise $\mathbf{e} = \bm{\epsilon}$ for one realization of a population $G_{U} = 2000$ and single two-stage cluster sample $G_S = 200$ under a symmetric design for both clusters and individuals.}
\label{fig:simprop}
\end{figure}

In Figure~\ref{fig:simprop}, we verify that the sample is `informative' and `balanced' with respect to the random effects and noise values. As expected, the population distribution of the random effects $\bm{a}$ is normally distributed. The PPS sampling leads to a bi-modal distribution of random effects in the sample, which in turn increases the variance of the sampled random effects relative to the population. The same effect is seen for the noise $\bm{\epsilon}$. Since the distributions are all mostly symmetric around the origin, the sampling design does not appear to be informative with respect to the intercept $b_0$. We note that the balance for $\epsilon$ satisfies our key condition \ref{itm:sampling design}\ref{withingroup}.

\subsubsection{Estimation Properties}
\begin{figure}
\centering
\includegraphics[width = 0.95\textwidth,
		page = 1,clip = true, trim = 0.0in 0.0in 0in 0.in]{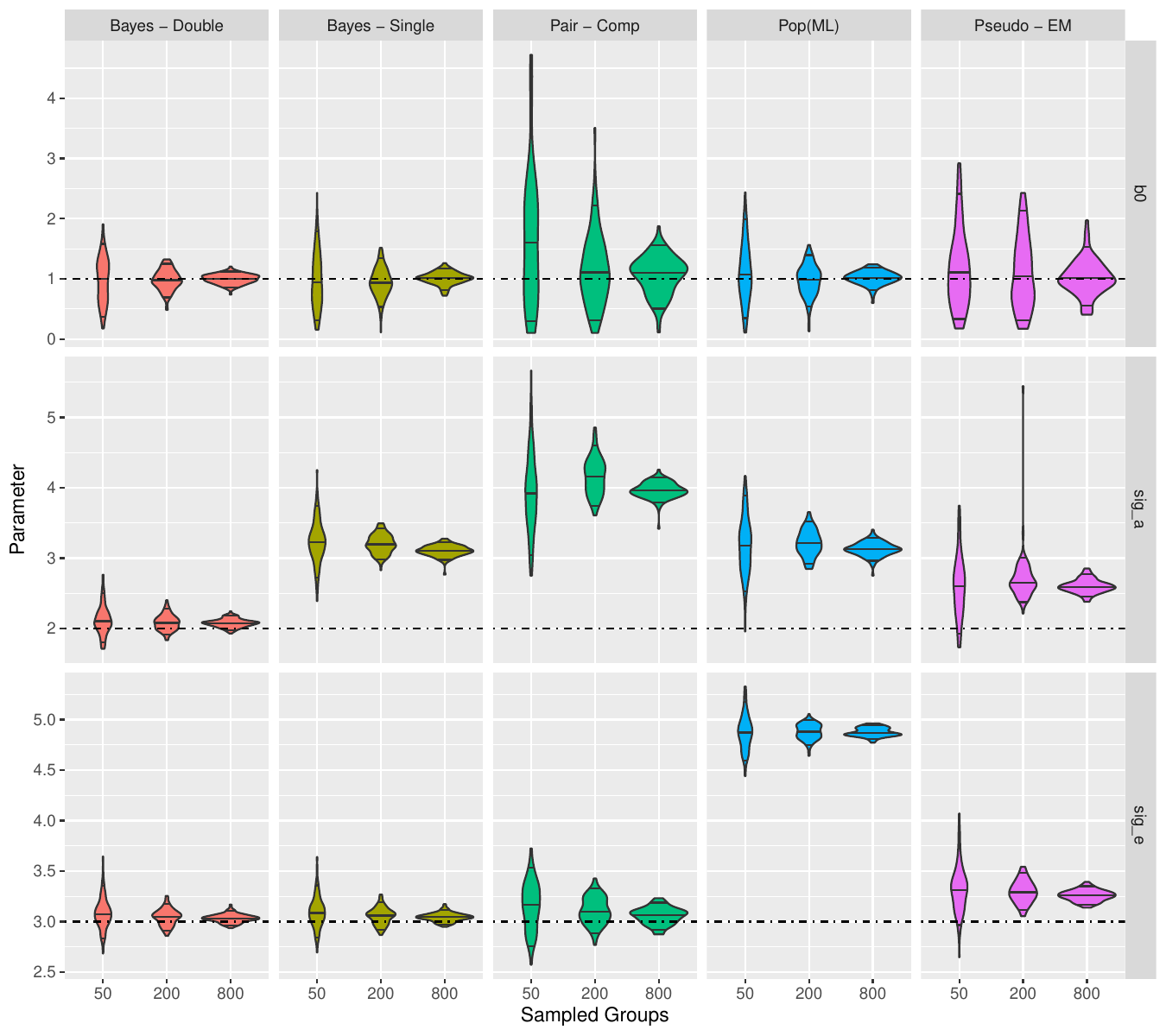}
\caption{Distribution and quantiles (5\%, 50\%, 95\%) of parameter estimates for R = 100 simulations by sample cluster size (x-axis), estimator (columns) and parameters (rows). Reference lines: population generating values.}
\label{fig:comparesim}
\end{figure}

In Figure~\ref{fig:comparesim}, we see that the naive use of equal weighting (Pop(ML)) leads to consistent bias for the estimates of both the random effects scale $\sigma_a$ and the noise scale $\sigma_\epsilon$. Including only the individual record level weights mitigates the bias for the noise scale $\sigma_\epsilon$, but appears to have no impact on the bias for the random effects scale $\sigma_a$. In contrast, using double weighting implemented via the augmented pseudo posterior leads to almost the complete removal of bias for the random effects scale. The pseudo EM method mitigates some bias for both $\sigma_a$ and  $\sigma_\epsilon$, but not as effectively as the double-weighting Bayes approach. (In addition the pseudo EM shows more variability between simulations). We include the pseudo EM of \citet{slud_2019} because it is a data augmentation estimator that is closest to our double weighted estimator.  Yet, \citet{slud_2019} requires \emph{non-informative} within cluster sampling in order to have a closed-form expectation for the E-step.  This restriction is likely why the pseudo EM performs not as well as the double weighted estimator.

Lastly, the pairwise composite approach mitigates much of the bias for $\sigma_{\epsilon}$ but demonstrates even more extreme bias for $\sigma_{a}$. In addition, the pairwise approach shows greater variability across replicates, which is expected from using a composite likelihood and second order weights.
Even for a moderate sample size of clusters (50), the differences are clear, with estimates showing contraction but minimal shifting with increases in the number of clusters sampled (200, and 800), while keeping the number of individuals sampled within-cluster fixed at 5.  As discussed in \citet{SavitskyWilliams+2022+901+928}, we believe that the integrating out of the random effects from the pairwise composite likelihood \emph{before} applying the cluster-indexed sampling disables the bias correction of the cluster sampling weights.

\begin{figure}
\centering
\includegraphics[width = 0.95\textwidth,
		page = 1,clip = true, trim = 0.0in 0.0in 0in 0.in]{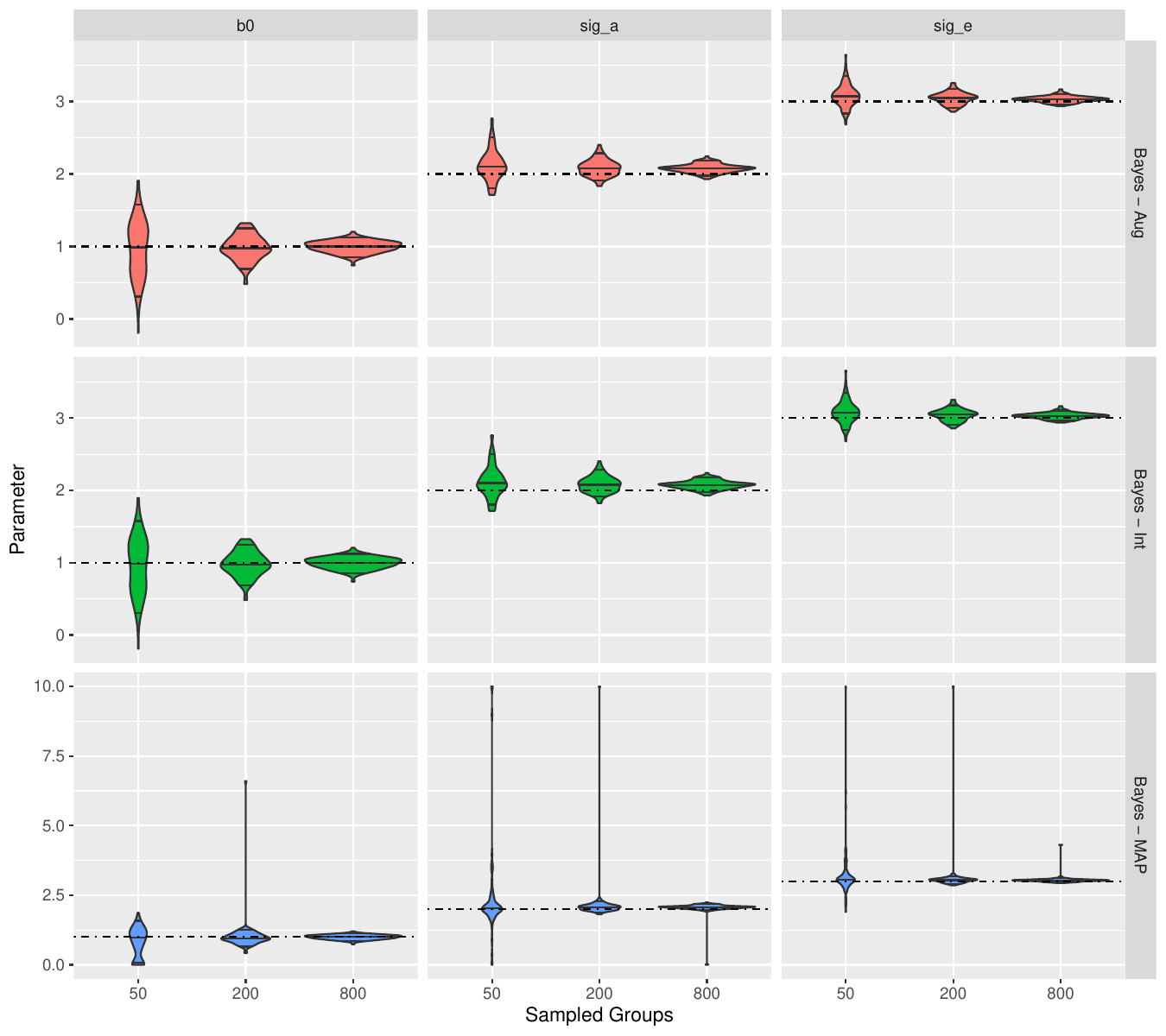}
\caption{Distribution and quantiles (5\%, 50\%, 95\%) of parameter estimates (cols) for R = 100 simulations for alternative estimation approaches (rows) for the double-weighted Bayesian approach for increasing sample cluster sizes (x-axis). Reference lines: population generating values. MAP Estimates truncated at 10.0 to preserve scale.}
\label{fig:comparedw}
\end{figure}

In Figure~\ref{fig:comparedw}, we next compare three alternative approaches to point estimation for the double-weighted pseudo Bayesian approach:
\begin{enumerate*}[label=(\roman*)]
\item MCMC under the augmented pseudo posterior of Equation~\ref{eq:sampcompletemodel} with generation of latent variables for random effects $a_{h}$,
\item MCMC using the integrated pseudo likelihood of Equation~\ref{eq:sampobsmodel} that marginalizes out $a_{h}$, and
\item the maximum a-posteriori (MAP) estimator using the integrated pseudo posterior under optimization.
\end{enumerate*} Both MCMC approaches lead to very similar results, as expected, with the main difference being in increased computational time for the augmented approach.  The integrated pseudo likelihood may be expressed as the product of survey-weighted normal distribution kernels, so we expect it to express the same asymptotic bias correction properties as the augmented pseudo likelihood.

The MAP estimator under the integrated pseudo posterior is less stable, but leads to similar point estimates, particularly for larger sample sizes. It also has the benefit of a significant reduction in computational time compared to the MCMC methods. Due to the weak prior information (which is asymptotically negligible), the MAP and the MLE using the integrated likelihood will be asymptotically equivalent.  Thus our general simulation results and the consistency conditions (Section \ref{sec:consistency}) of the augmented and integrated likelihood also apply to the double-weighted pseudo MLE under the integrated likelihood. However, as we see in our simulations, stability of argmax estimators may be an issue.

\section{Simulation Study 2: Exploring Consistency Requirements}\label{sec:sims2}
The second simulation study modifies the sampling designs above to examine cases where consistent results are not necessarily expected. In particular, we look at unbalanced \emph{asymmetric} sampling designs at one or both stages that induce highly skewed samples. As demonstrated in Section \ref{sec:consistency}, asymmetric designs may violate a condition in balance for within-cluster weighted residuals needed to guarantee consistency for arbitrarily small within-cluster sampling fractions to varying degrees; however, even for these extreme designs, we see the improvement to estimation from moderate increases in the within cluster sampling fraction, which aligns with the theory. Furthermore, it is valuable to compare the alternative estimators and to gauge the robustness of competing approaches. In practice, the true informativeness of the design is difficult to know ($\epsilon_{h\ell}$ are not usually observed during the sampling stage), so more robust estimation approaches are preferred.

\subsection{Sample Design and Estimation}\label{sec:sim2des}
The model is the same as in Section \ref{sec:sims1model}. The sample design is similar to that of Section \ref{sec:sims1samp}, however we modify the PPS size measures:

\begin{enumerate}
\item Generate R = 100 populations from the one-way ANOVA distribution (using true values, $\mu_{0} = 1$, $\tau_{a}^{-1} = 2$, $\tau_{\epsilon}^{-1} = 3$).
	\begin{enumerate}
		\item With $M = \{2000\}$ clusters
		\item Each with $N_{h} = \{40\}$ individuals in each cluster
	\end{enumerate}
\item For each $r = 1,\ldots, R$ population, draw a two-stage sample via `mstage'  in R \citep{sampling}
	\begin{enumerate}
		\item Sample $m = \{200\}$ clusters. Using systematic PPS sampling with size:
		\begin{enumerate}
			\item  $\pi_h \propto (a_{h})^2 + 1$ (``quadratic symmetric'') or
			\item $\pi_h	\propto a_{h} - \min_{h}(a_{h}) + 1$ 	(``linear asymmetric'').
		\end{enumerate}
		\item Sample $n_{k} = \{5,10,20\}$ individuals in each cluster design. Using systematic PPS sampling with size
		\begin{enumerate}
			\item  $\pi_{\ell|h} \propto \max(0,\epsilon_{h \ell})^2 + 1$ (``quadratic'')
			\item  $\pi_{\ell|h} \propto 0.3 \max(0,\epsilon_{h \ell})^2 + 1$ (``weak quadratic'')
			\item  $\pi_{\ell|h} \propto \epsilon_{h \ell} - \min_{h \ell}+ 1$ (``linear'')
			\item  $\pi_{\ell|h} \propto 0.3 (\epsilon_{h \ell} - \min_{h \ell})+ 1$ (``weak linear'')
			\item  $\pi_{\ell|h} \propto 1$ (``simple random sample'')
		\end{enumerate}
	\end{enumerate}
\item For each $r = 1,\ldots, R$ sample, estimate $\{b_0, \sigma_a, \sigma_{\epsilon}\}$
	\begin{enumerate}
		\item Using equal weights $w_{k} = 1$,  $w_{k j} = 1$ via `lmer' in R \citep{lme4}
		\item Using single weights $w_{k} = 1$,  $w_{k j} \propto 1/\pi_{k j}$ via Stan \citep{stan:2015}
		\item Using double weights $w_{k} \propto 1/\pi_{k}$,  $w_{k j} \propto 1/\pi_{k j}$ via Stan \citep{stan:2015}.
		\item Using double weights $w_{k} \propto 1/\pi_{k}$,  $w_{k j} \propto 1/\pi_{k j}$ in an EM algorithm as in \citet{slud_2019}
		\item Using two stage weights $w_{k} \propto 1/\pi_{k}$,  $w_{j|k} \propto 1/\pi_{j|k}$ and the pairwise composite method \citep{yi:2016} as implemented in `svylme' in R \citep{svylme}.
	\end{enumerate}
\end{enumerate}

\subsection{Results}
\subsubsection{Sample Properties}

\begin{figure}
\centering
\includegraphics[width = 0.50\textwidth,
		page = 1,clip = true, trim = 0.0in 0.0in 0in 0.in]{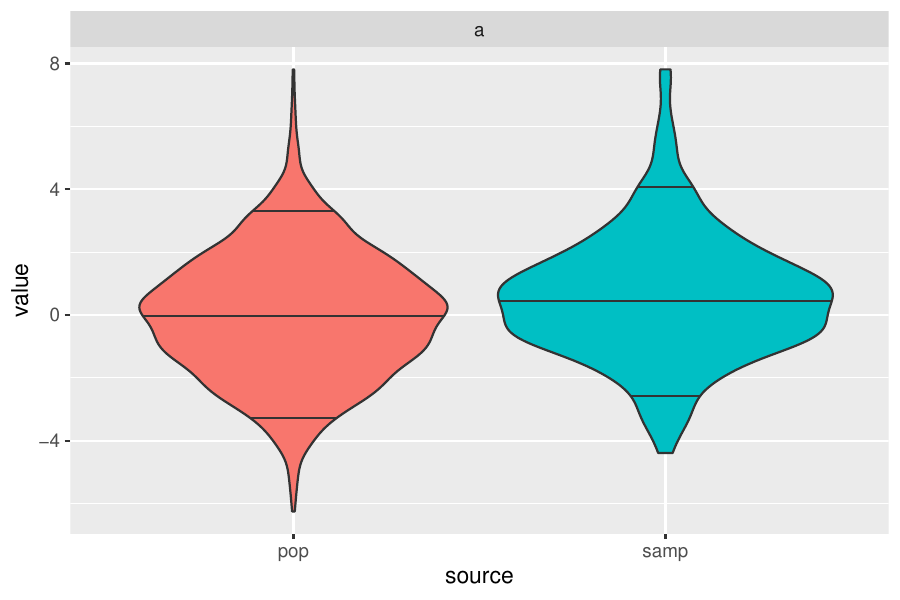}
\caption{Distributions and quantiles (5\%, 50\%, 95\%) of random effects $\mathbf{a}$ for one realization of a population $G_{U} = 2000$ and single two-stage cluster sample $G_S = 200$ under asymmetric sampling design for clusters}.
\label{fig:simprop_asymm_a}
\end{figure}

In Figure \ref{fig:simprop_asymm_a} we see that the linear asymmetric design for sampling clusters leads to a shifted and skewed sample of $a_k$ relative to the population. Unlike the symmetric design for clusters (Figure \ref{fig:simprop}), this shift and skew means that the estimate for the intercept $b_0$ will be impacted. The estimate for $\sigma_a$ will be impacted as well, but to a lesser extent than for the symmetric quadratic design.

\begin{figure}
\centering
\includegraphics[width = 0.95\textwidth,
		page = 1,clip = true, trim = 0.0in 0.0in 0in 0.in]{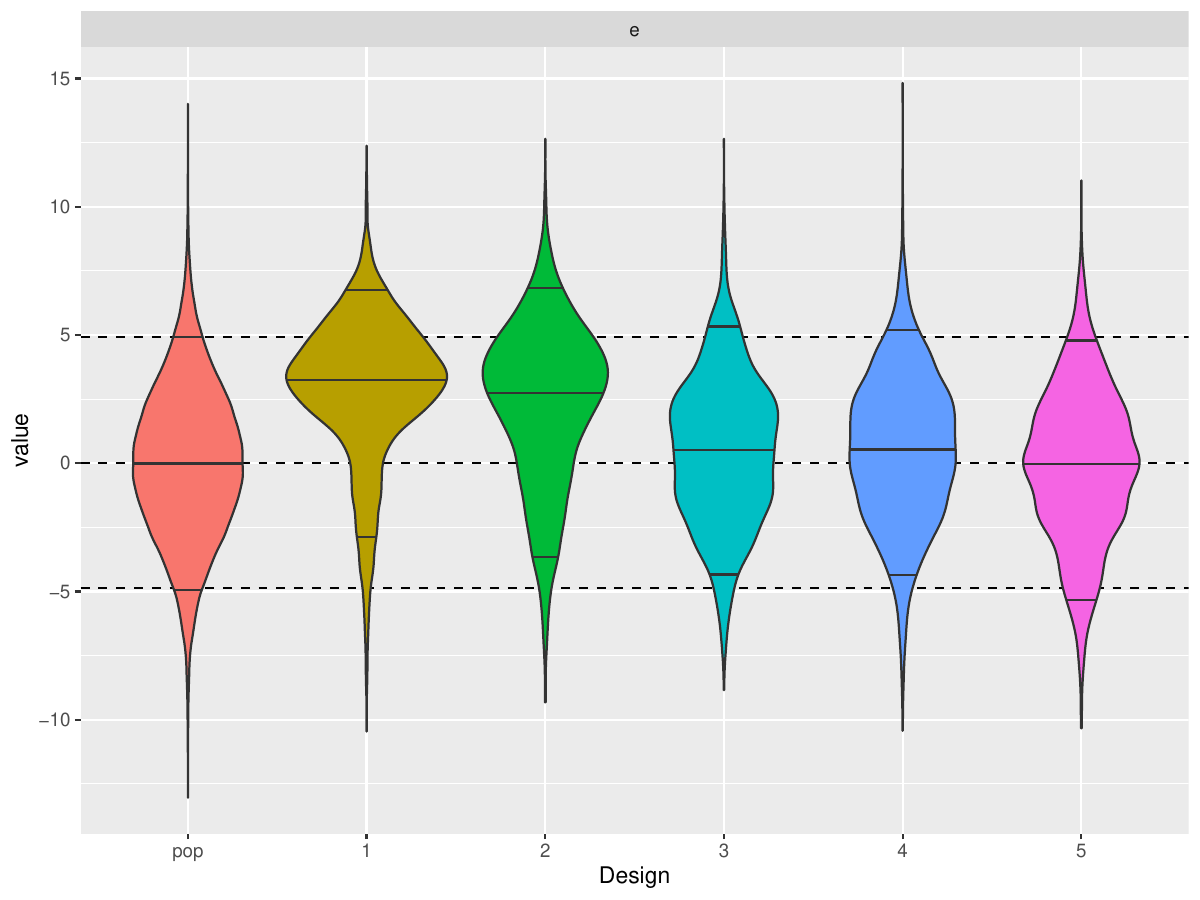}
\caption{Distributions and quantiles (5\%, 50\%, 95\%) of noise $\mathbf{e} = \bm{\epsilon}$ for one realization of a population $G_{U} = 2000$ and two-stage cluster samples under alternative with-in cluster designs with $m = 200$ clusters sampled and $n_k = 10$ individuals within each cluster sampled. Quadratic (1), Weak Quadratic (2), Linear (3), Weak Linear (4), SRS (5). }
\label{fig:simprop_asymm}
\end{figure}

Figure \ref{fig:simprop_asymm}, compares the population distribution of $\epsilon_{\ell h}$ to that from within cluster sampling designs for four unbalanced asymmetric designs and an SRS for $n_k = 10$. The quadratic asymmetric designs have an extreme shift and skew, with few sampled $\epsilon_{jk}$ that are negative. In other words, many observed samples will have some clusters with only positive $\epsilon_{jk}$. The linear asymmetric designs still display a shift and skew, but have a much more balanced proportion of positive and negative $\epsilon_{jk}$. The shifted and skewed distributions for sampled $\epsilon_{jk}$ are informative with respect to both the estimates of the intercept $b_0$ and the variance parameter $\sigma_{\epsilon}$. We expect estimates for $\sigma_{\epsilon}$ will be impacted to a lesser extent than for the symmetric quadratic design (Figure \ref{fig:simprop}).

\subsubsection{Estimation Properties}
\begin{figure}
\centering
\includegraphics[width = 0.95\textwidth,
		page = 1,clip = true, trim = 0.0in 0.0in 0in 0.in]{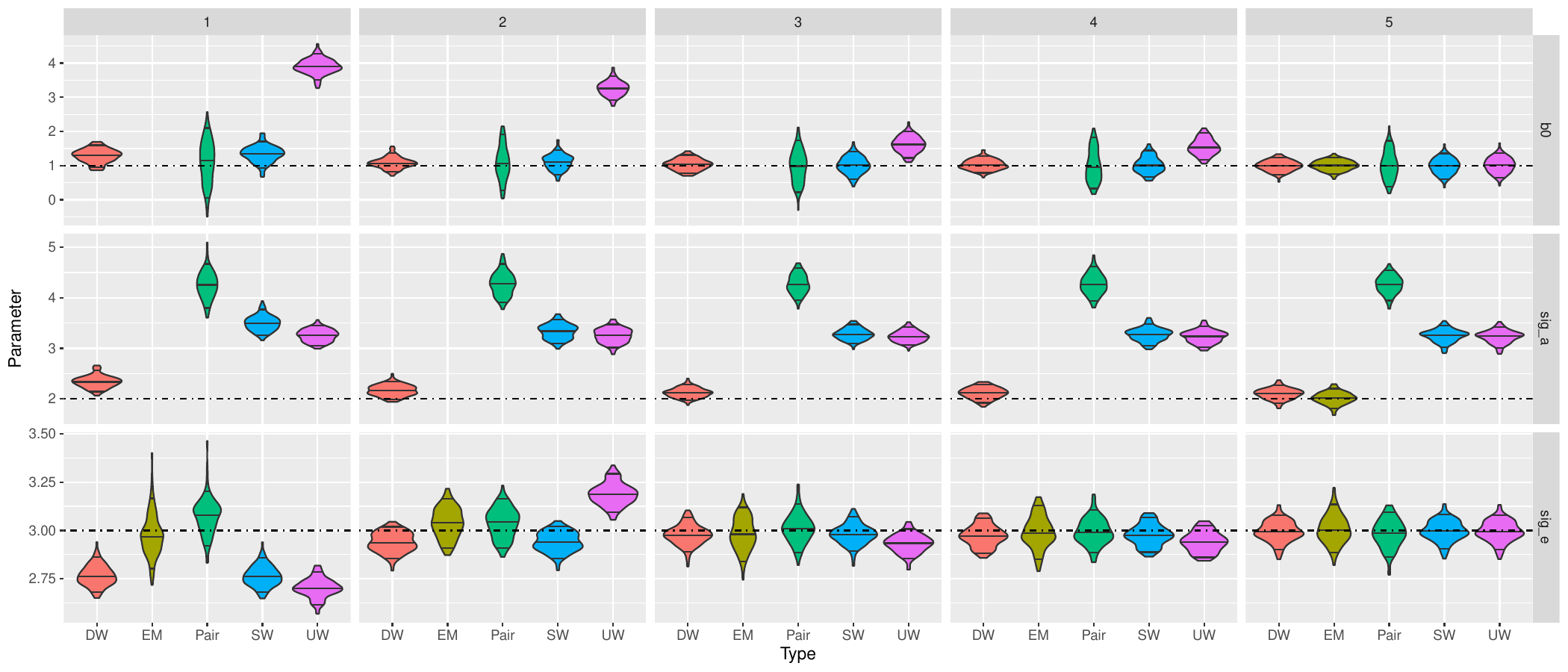}
\caption{Distributions and quantiles (5\%, 50\%, 95\%) of parameter estimates for R = 100 simulations for estimator (x-axis) and parameters (rows) across for \emph{symmetric quadratic first stage} sample with varying second stage sample designs (cols) for sample sizes $n_k = 10$: Quadratic (1), Weak Quadratic (2), Linear (3), Weak Linear (4), SRS (5). Reference lines: population generating values.}
\label{fig:compare_symclust}
\end{figure}

\begin{figure}
\centering
\includegraphics[width = 0.95\textwidth,
		page = 1,clip = true, trim = 0.0in 0.0in 0in 0.in]{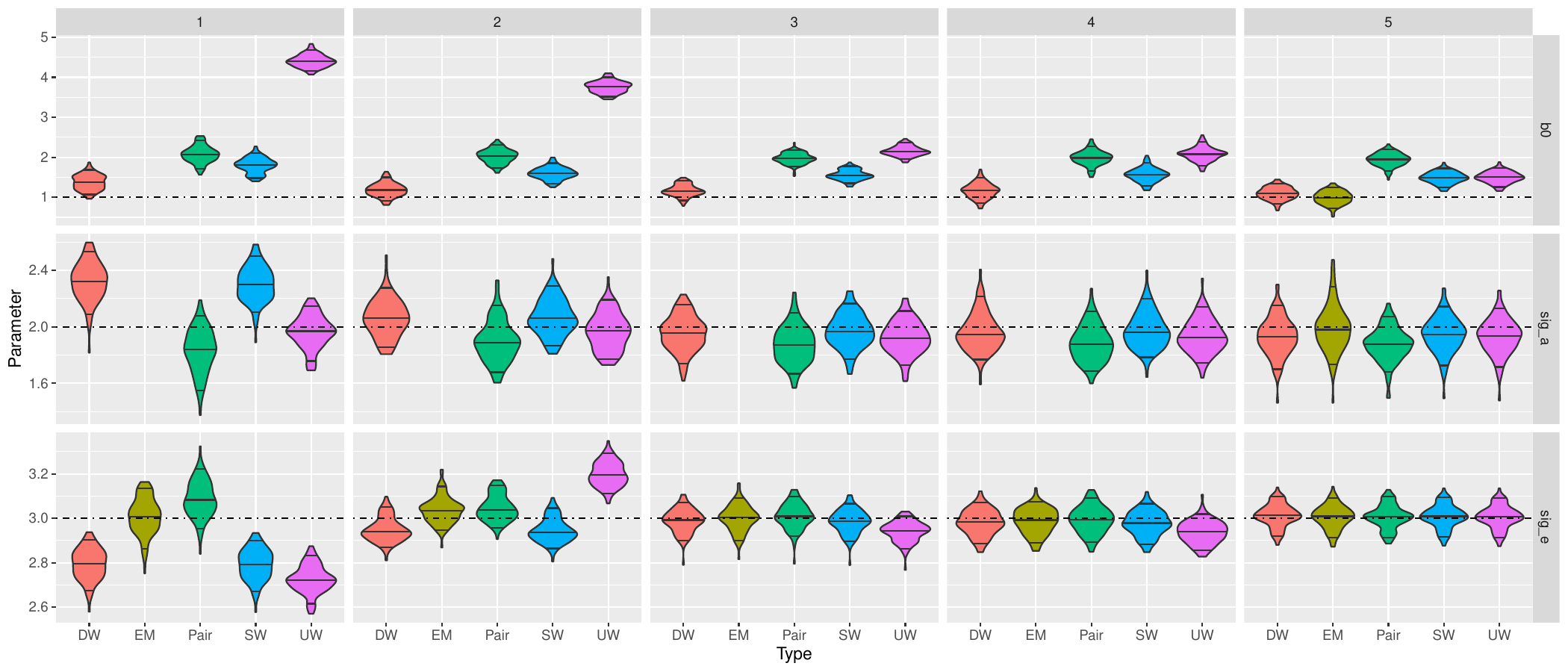}
\caption{Distributions and quantiles (5\%, 50\%, 95\%) of parameter estimates for R = 100 simulations for estimator (x-axis) and parameters (rows) across for \emph{asymmetric linear first stage} sample with varying second stage sample designs (cols) for sample sizes $n_k = 10$: Quadratic (1), Weak Quadratic (2), Linear (3), Weak Linear (4), SRS (5). Reference lines: population generating values.}
\label{fig:compare_asymclust}
\end{figure}

Figure \ref{fig:compare_symclust} displays the distribution of parameter estimates across $R = 100$ replications of the symmetric between cluster sampling and the 5 variations of within cluster sampling for $n_k = 10$. Figure \ref{fig:compare_asymclust} displays the analogous results for asymmetric between cluster sampling. We summarize results by the estimation method. General observations are similar for $n_k =5, 20$ (results not included).

\begin{enumerate}
	\item[{\bf Unweighted}] The unweighted MLE estimation provides a baseline to demonstrate the biases induced from informative sampling, which the other methods aim to mitigate. For the symmetric cluster sampling design (Figure \ref{fig:compare_symclust}) there is a clear expected upward bias for the random effects variance $\sigma_a$. For asymmetric sampling of residuals, we see an upward bias of the intercept $b_0$. There is some bias or instability of the estimate of $\sigma_{\epsilon}$ but we see that the magnitude is significantly smaller than the biases for the $b_0$ and $\sigma_a$. For the asymmetric between cluster designs (Figure \ref{fig:compare_asymclust}) we see an expected increase in the bias for $b_0$ and a reduced bias for $\sigma_{a}$ due to the skewed sampling of $a_{h}$ (Figure \ref{fig:simprop_asymm_a}).
	\item[{\bf Single Weighted}] The single weighted approach only adjusts the likelihood contributions of individuals but leaves the group level random effects distribution un-adjusted. The main improvement over the unweighted approach is the reduction in bias for the intercept $b_0$ along with some improved stability for estimation of $\sigma_{\epsilon}$; however, biases from the informative sampling between clusters (bias with respect to $\sigma_{a}$ for symmetric designs, Figure \ref{fig:compare_symclust} and bias with respect to $b_0$ for asymmetric designs, Figure \ref{fig:compare_asymclust}) still remain.
	\item[{\bf Double Weighted}] The double weighted method largely mitigates the biases demonstrated by the unweighted estimator,  including the remaining bias demonstrated by the single weighted estimator. In particular, the large upward bias for $\sigma_{a}$ for symmetric between clustering designs is mostly mitigated as well as the large upward bias for $b_0$ for asymmetric between clustering designs.  The major area of concern is for the extremely skewed informative within cluster sampling design (1).  We see in Figure~\ref{fig:simprop_asymm} that when the within clusters residuals are highly unbalanced, the weighted residuals are also far from balanced (not centered at $0$) as required by Condition~\ref{itm:sampling design}\ref{withingroup} in Section~\ref{sec:consistency}, as is strongly the case in design (1), such that double weighting will not be consistent for $\sigma_a$, as is revealed in Figures~\ref{fig:compare_symclust} and \ref{fig:compare_asymclust}; however with increasing (within cluster) sample size (Section \ref{sec:sims_consist}) and slightly less informative designs (2,3), the double-weighted method demonstrates less bias compared to the unweighted and single-weighted designs.  As the theoretical condition for consistency of $\sigma_a$ is met where balance is achieved in the weighted within-cluster residuals for any of the sampling designs, we see in the figures that $\sigma_a$ contracts on the truth.
	\item[{\bf EM}] In the SRS within-cluster designs, the EM method shows essentially no bias for all three parameter estimates for both symmetric and asymmetric between cluster designs. This is predicted by the theory developed in \citet{slud_2019} that demonstrates consistency of the EM under non-informative within-cluster sampling. Unfortunately, the EM method shows extreme instability for even weakly informative within-cluster designs. No estimates for $b_0$ or $\sigma_{a}$ are in the interval [-10,10], with many much farther out. Even with unusable estimates for intercept and random effects variance, the estimates for $\sigma_{\epsilon}$ appear to be more robust.
	\item[{\bf Pairwise}] In contrast to the EM estimator, the pairwise estimator demonstrates significant biases even when the within cluster sampling is SRS.  For the symmetric cluster sampling designs (Figure \ref{fig:compare_symclust}), this manifests as a large upward bias for $\sigma_{a}$. The pairwise estimator completely breaks down for estimation of $\sigma_a$ under symmetric sampling of clusters as the pairwise weights fail to correct the upward bias in the sample induced by an informative symmetric first stage design. Bias in $\sigma_a$ is persistent, though relatively small, even for the asymmetric first stage designs (Figure \ref{fig:compare_asymclust}), though unlike the case of the double weighted estimator, this bias is insensitive to residual balance within clusters. In effect, the pairwise estimator fails to correct for the informative sampling of random effects because the estimator of \citet{yi:2016} marginalizes out the random effects, $\mathbf{a}$, from the joint likelihood for each nested pair of observations before applying group weight, $w_{k}$, such that the sample-based informativeness between clusters is not moderated or adjusted by the first stage weights.  Similar to the EM, across both sets of designs, the estimate for $\sigma_{\epsilon}$ appears more stable. We contrast that with the symmetric within cluster designs (Figure \ref{fig:comparesim}) in which both the pairwise and EM show some bias for $\sigma_{\epsilon}$. We also note that the relative bias for $\sigma_{\epsilon}$ across all estimators is small for these designs.
\end{enumerate}

\subsubsection{Consistency of the Double-Weighting Approach}\label{sec:sims_consist}
\begin{figure}
\centering
\includegraphics[width = 0.95\textwidth,
		page = 1,clip = true, trim = 0.0in 0.0in 0in 0.in]{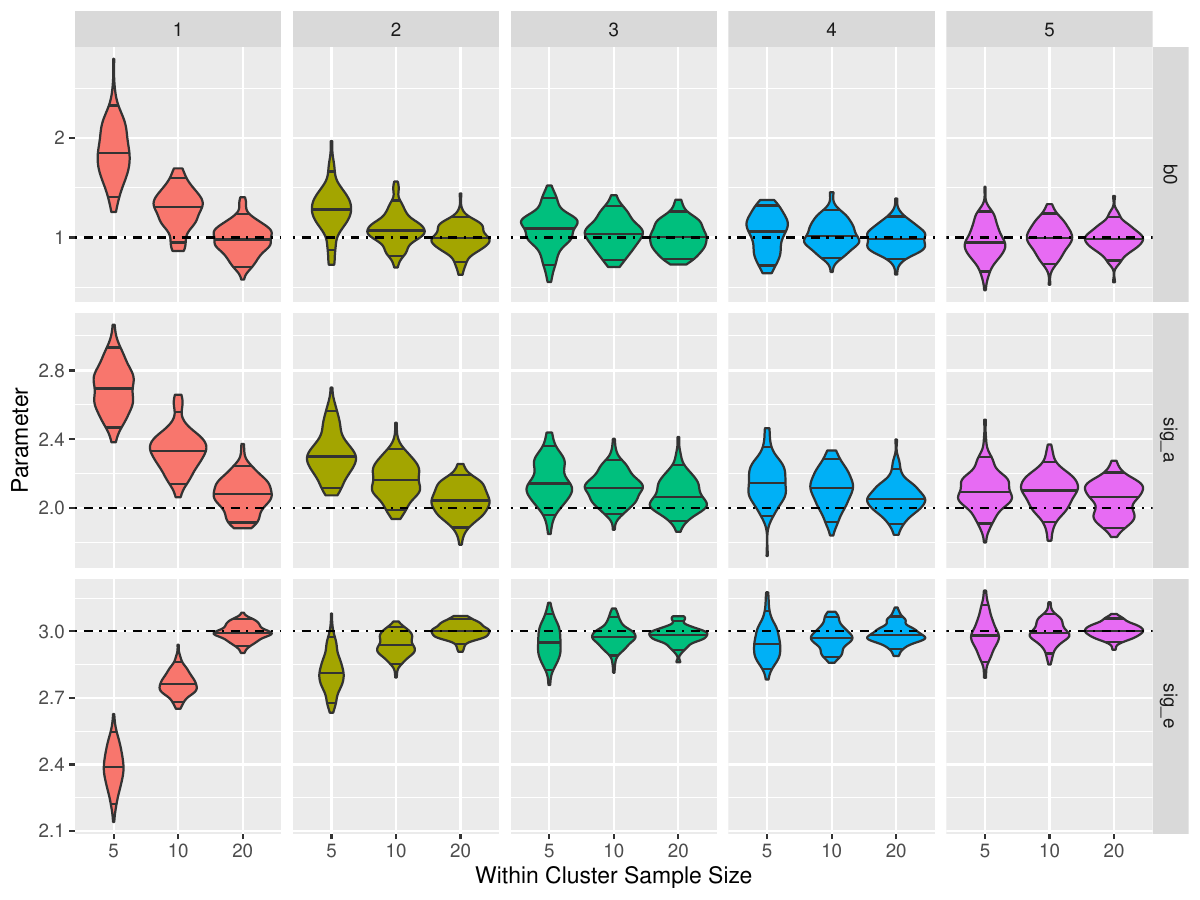}
\caption{Distributions and quantiles (5\%, 50\%, 95\%) of parameter estimates for R = 100 simulations for the double-weighting estimator by within cluster sample size (x-axis) and parameters (rows) across for \emph{symmetric quadratic first stage} sample with varying second stage sample designs (cols): Quadratic (1), Weak Quadratic (2), Linear (3), Weak Linear (4), SRS (5). Reference lines: population generating values.}
\label{fig:dw_symclust_nk}
\end{figure}

\begin{figure}
\centering
\includegraphics[width = 0.95\textwidth,
		page = 1,clip = true, trim = 0.0in 0.0in 0in 0.in]{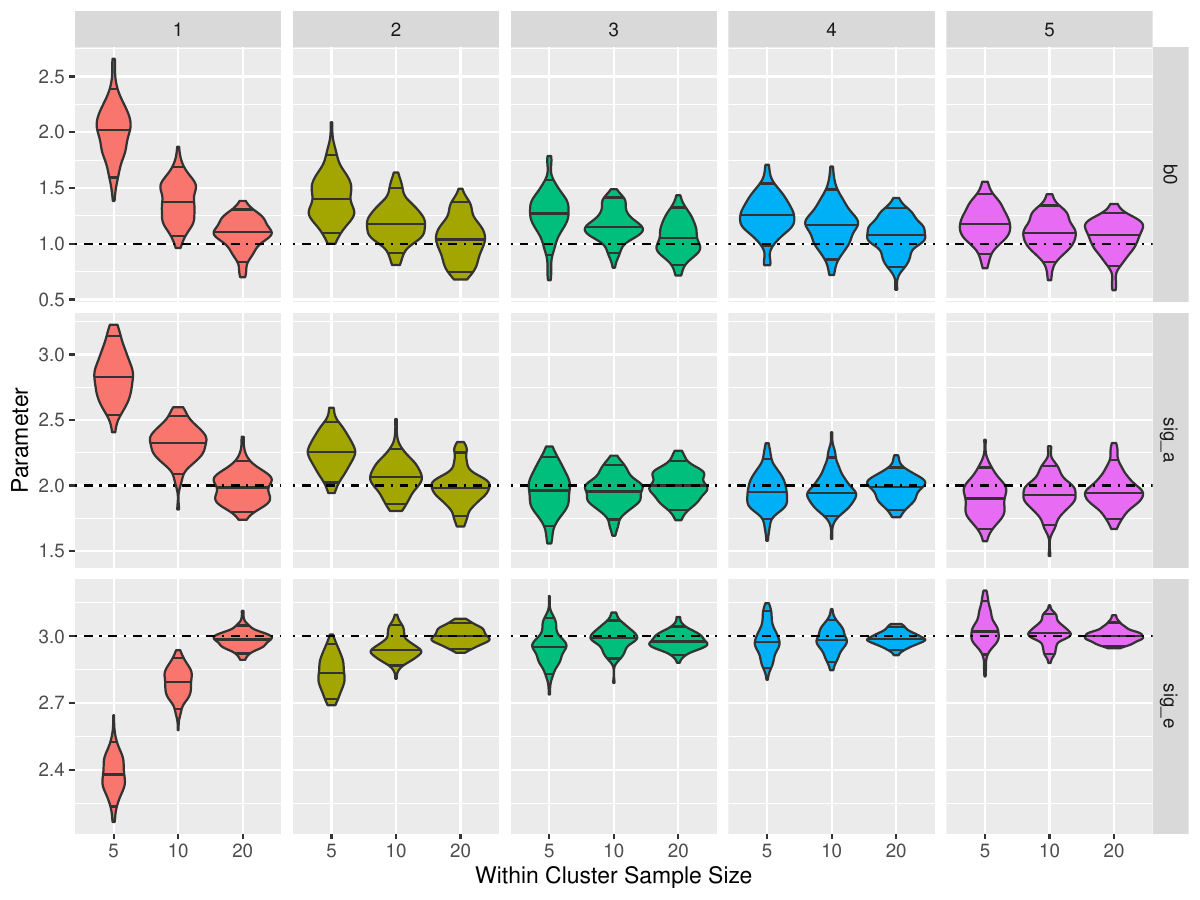}
\caption{Distributions and quantiles (5\%, 50\%, 95\%) of parameter estimates for R = 100 simulations for the double-weighting estimator by within cluster sample size (x-axis) and parameters (rows) across for \emph{asymmetric linear first stage} sample with varying second stage sample designs (cols) Quadratic (1), Weak Quadratic (2), Linear (3), Weak Linear (4), SRS (5). Reference lines: population generating values.}
\label{fig:dw_asymclust_nk}
\end{figure}

In Figures \ref{fig:dw_symclust_nk} and \ref{fig:dw_asymclust_nk}, we show the behavior of the double-weighting approach as the sampling fraction within each cluster increases from $1/8$ ($n_k = 5$) to $1/2$ ($n_k = 20$) across all asymmetric sampling designs. For the extreme designs (asymmetric quadratic) a significant improvement occurs between $n_k = 10$ and $n_k=20$.
For moderately informative designs (asymmetric linear) the sample size of $n_k = 10$ seems about as effective as $n_k=20$. For the uniformative within cluster design (SRS), the smallest sample size $n_k =5$ has about the same bias as the larger sample sizes.
This is consistent with the theory in Section \ref{sec:consistency} which suggests that more strongly informative within-cluster designs will need a larger sampling fraction to obtain consistent estimation.
One main difference between Figures \ref{fig:dw_symclust_nk} and \ref{fig:dw_asymclust_nk} is the small but decernable residual bias for $\sigma_a$ in Figure \ref{fig:dw_symclust_nk}. This is likely caused by the very strong symmetric sampling design for the first stage clusters, which tends to skew the estimates of $\sigma_a$. We expect this bias to decrease further for increasing number of sampled clusters (e.g. increasing $m > 200$).

\FloatBarrier

\section{Application: The PISA 2000}
The OECD Programme for International Student Assessment (PISA) 2000 compiles student and school data in 32 countries \citep{organisation2000database}. In each country, schools are sampled proportional to enrollment. Within schools, students are sampled with equal probability. Based on individual student assessments, multiple plausible values are imputed for reading, mathematics, and science. These plausible values are based on a transformed index and are strongly normally distributed. For illustration, we subset to the 153 schools and 2129 students from the United States. For a response variable, we choose the average of the 5 imputed plausible values for science. This allows us to directly implement the two-level or one-way ANOVA model developed in the theory and simulations. We note that adding random effects for country and student repeated measures would allow for a richer four-level analysis of variance. Such is readily encompassed in the general approach developed in this paper but is beyond the scope of this exposition.

\begin{figure}
\centering
\includegraphics[width = 0.95\textwidth,
		page = 1,clip = true, trim = 0.0in 0.0in 0in 0.in]{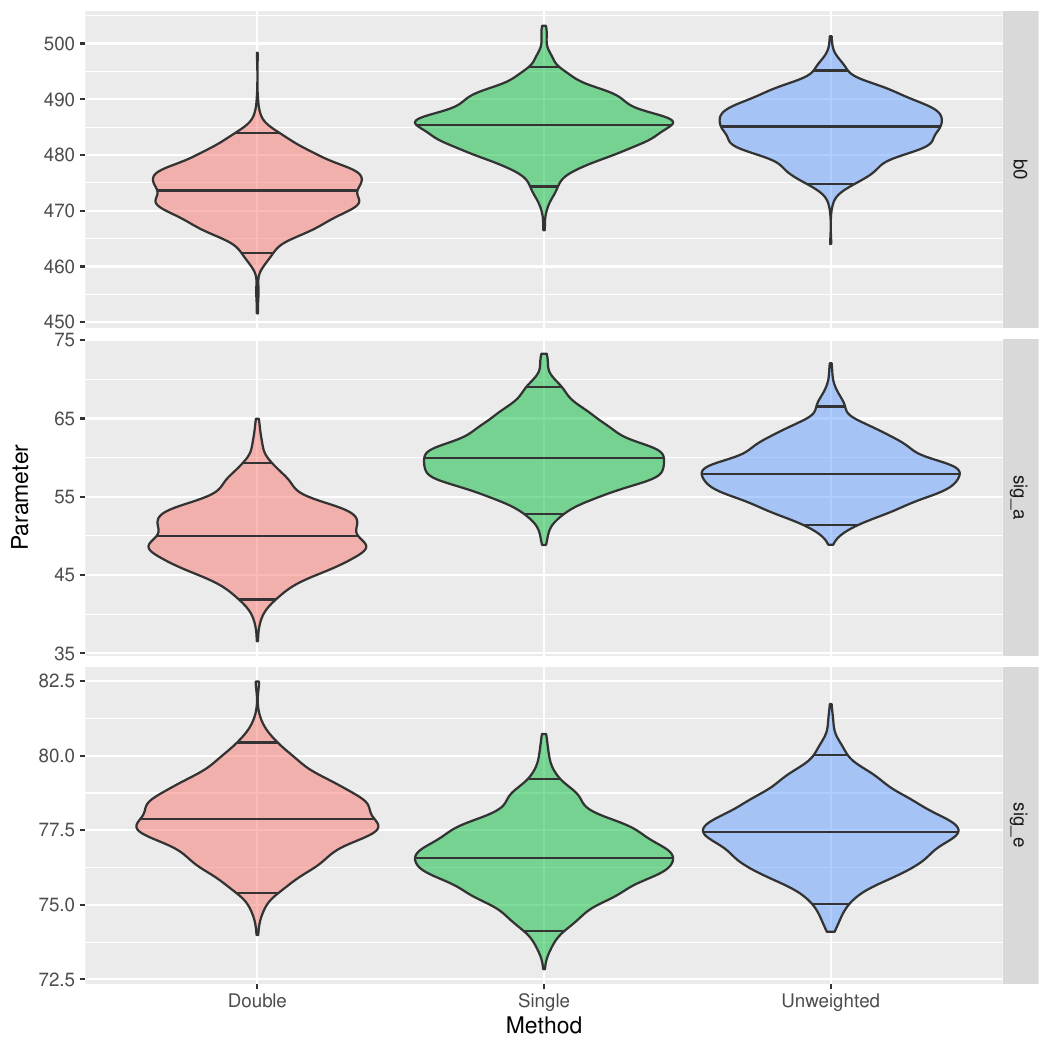}
\caption{Distribution and quantiles (5\%, 50\%, 95\%) of parameter estimates for ANOVA of science scores for students in US schools from PISA 2000, Posterior estimator (x-axis) and parameters (rows).}
\label{fig:PISA compare}
\end{figure}

Figure \ref{fig:PISA compare} compares the three model parameters (intercept, school variability, individual variability) estimated by unweighted, single-weighted, and double-weighted pseudo Bayesian methods. There are only minor differences between the unweighted and single-weighted results - with the single-weighted leading to slightly smaller values for $\sigma_a$ and slightly larger values for $\sigma_{\epsilon}$. The double weighting approach leads to larger differences in both the variance components and the intercept. As shown in the simulations, when designs are not orthogonal between first and second stage sampling, bias estimating one of the three parameters impacts bias for one or both of the other parameters. Based on our theory and simulation study, we would expect that the double-weighting correctly adjusts for the two-phase sampling design, while unweighted and single-weighting have inflated the intercept and over-estimated the contribution of school to the variance, while underestimating the variability between individual students.

\begin{figure}
\centering
\includegraphics[width = 0.95\textwidth,
		page = 1,clip = true, trim = 0.0in 0.0in 0in 0.in]{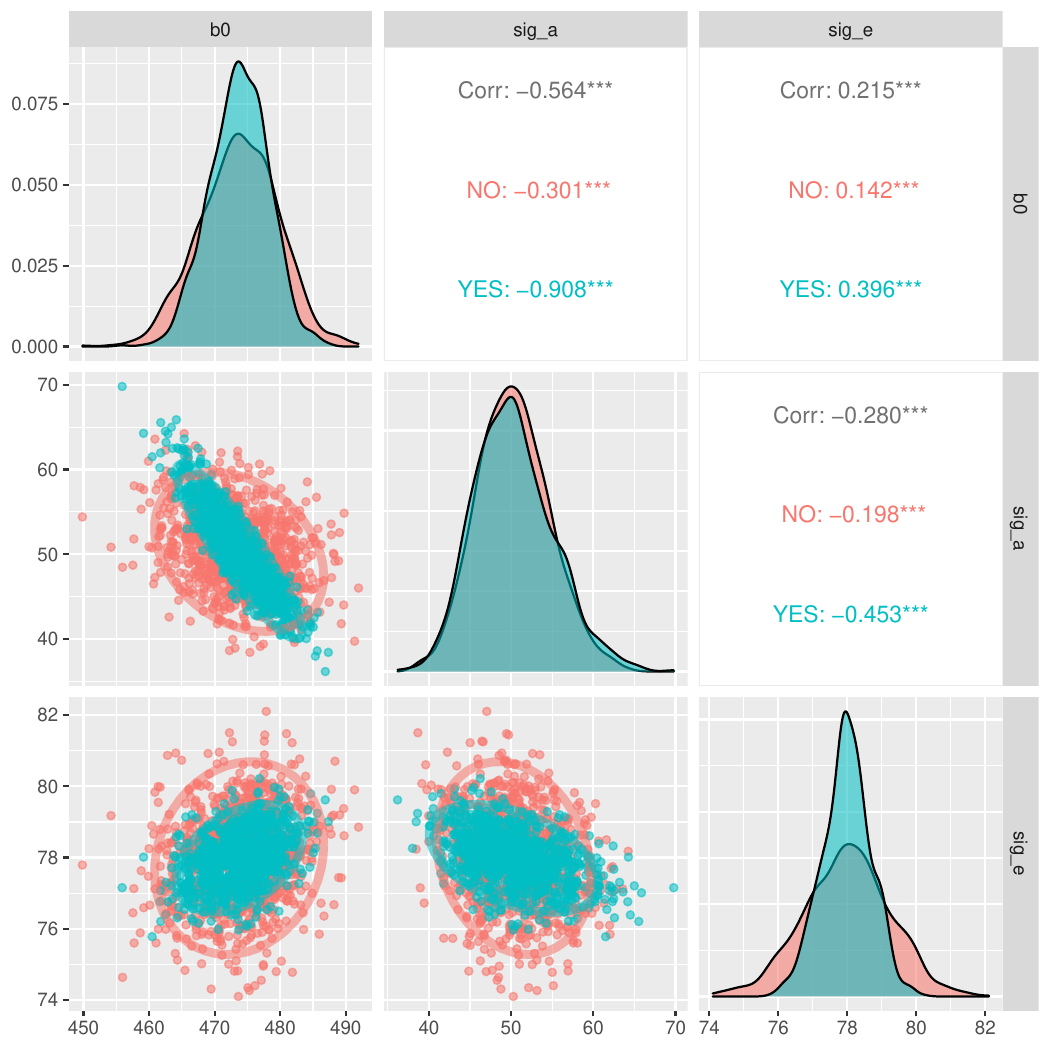}
\caption{Comparison of double-weighted posterior sample before (NO/Orange) and after (YES/Blue) covariance adjustment. Marginal density plots (on diagonal), bivariate scatter plots with approximate 90\% ellipse (below diagonal), and Pearson correlation (above diagonal).}
\label{fig:PISA adjust}
\end{figure}

To investigate the uncertainty quantification of the double-weighting method, we employ the approach of \cite{WilliamsISR2021} to estimate an asymptotic sandwich adjustment for the posterior variance of the global parameters. Figure \ref{fig:PISA adjust} shows that after properly modelling the school-level contribution, the marginal variability for both the intercept and the student-level variance are conservative (too large) if left unadjusted (orange). We also note that the design-induced correlation between the intercept and the school-level variance is not properly captured without adjustment (such a correlation is not parameterized in a simple one-way ANOVA model). This aligns well with our simulation study results, in which we observed strong correlation between estimates for the intercept and cluster-level random effects variance component.

\section{Discussion}
We have used the canonical one-way ANOVA mixed effects model to enumerate conditions that guarantee the consistency of a sample-based estimator that incorporates marginal sampling weights under the two-stage informative sampling of clusters and units within clusters. 
A common motivating example is the assessment of students nested within schools.  
Our ``double-weighted" estimator creates an augmented pseudo posterior formulation that is estimated on the observed sample by exponentiating the prior for latent random effects, $\{a_{k}\}$, by the marginal cluster sampling weights, $\{w_{k}\}$, in addition to the usual procedure of exponentiating the observed data likelihood for, $\{y_{jk}\}$, by the marginal unit sampling weights, $\{w_{jk}\}$.

Our theoretical results reveal a class of balanced (`symmetric') and weakly to moderately unbalanced (`asymmetric') sampling designs for which frequentist consistency of the marginally-weighted, double-weighted estimator is guaranteed, as well as class of highly unbalanced sampling designs for which consistency would not be expected to be achieved.  The class of sampling designs under which consistency of our double-weighted estimator is guaranteed encompasses many sampling designs used in practice, making our consistency result to be highly relevant since the double-weighted estimator employs only marginal sampling weights.

\FloatBarrier
\vskip 0.2in
\bibliography{refs_april_2022}
\bibliographystyle{apalike}

\newpage

\appendix


\section{Proof of Proposition \ref{prop1}}
\label{sec:proof-prop-refpr}

Equations \ref{ak-post-1} and \ref{ak-post-2} imply that
\begin{equation}
  \left(a_{k}\middle\vert\{y_{jk}\}, \{w_{jk}\} , \{w_{k}\}, \tau_{\epsilon}, \tau_{a}\right) = \mathcal{N}\left(a_{k}\middle\vert h_{k},\phi_{k}^{-1}\right),
\end{equation}
where $e_{k} = \tau_{\epsilon}\mathop{\sum}_{j\in S_{k}}w_{jk} (a_{k0} + \epsilon_{jk0})$, $\phi_{k} = \tau_{\epsilon}\mathop{\sum}_{j\in S_{k}}w_{jk}\ + \tau_{a}w_{k}$ and $h_{k} = \phi_{k}^{-1}e_{k}$. Equation
\ref{eq:wtd-avg} implies that the weighted average of random effects satisfies
\begin{equation}\label{expas}
\begin{array}{l}
\mathbb{E}\left (M^{-1}\sum_{k \in S_{c}} w_{k} a_{k}\middle\vert\{y_{jk}\}, \{w_{jk}\} , \{w_{k}\}, \tau_{\epsilon}, \tau_{a}\right)  \\
\quad  = M^{-1} \sum _{k \in S_{c}} w_{k} \left( \dfrac{\tau_\epsilon  \sum_{j \in S_{k}} w_{jk}(a_{k0} + \epsilon_{jk0})}
	{\tau_\epsilon \sum_{j \in S_{k}} w_{jk} + \tau_{a}  w_{k}} \right)\\
\quad =   M^{-1} \sum _{k \in S_{c}} w_{k} a_{k0} \left( \dfrac{\sum_{j \in S_{k}} w_{j|k}}
	{\sum_{j \in S_{k}} w_{j|k} + \tau_{a}/\tau_{\epsilon}} \right) +  \	M^{-1} \sum _{k \in S_{c}} w_{k} \left( \dfrac{\sum_{j \in S_{k}} w_{j|k}\epsilon_{jk0}}
	{\sum_{j \in S_{k}} w_{j|k} + \tau_{a}/\tau_{\epsilon}} \right) \\
\quad =  M^{-1} \sum_{k \in S_{c}} w_{k} a_{k0} \left( \dfrac{\hat{N}_{k}}
	{\hat{N}_{k} + \tau_{a}/\tau_{\epsilon}} \right) +  \	M^{-1} \sum _{k \in S_{c}} w_{k} \left( \dfrac{\sum_{j \in S_{k}} w_{j|k}\epsilon_{jk0}}
	{\hat{N}_{k} + \tau_{a}/\tau_{\epsilon}} \right).
\end{array}
\end{equation}
We expand the last line in Equation \ref{expas} from the observed sample to the population of clusters by inserting group-level inclusion indicators, $(\delta_{h})_{h=1}^{M}$, in
\begin{equation}
\begin{array}{rl}
B^{w} \ = & M^{-1}\sum_{h=1}^{M} \mathbbm{1} \left\{h \in S_{c}\right\} w_{h} \left\{a_{h0} \left( \dfrac{\hat{N}_{h}}
	{\hat{N}_{h} + \tau_{a}/\tau_{\epsilon}} \right) +  \left( \dfrac{\sum_{j\in S_{h}} w_{j|h}\epsilon_{j h0}}
	{\hat{N}_{h} + \tau_{a}/\tau_{\epsilon}} \right) \right\}
\end{array}
\end{equation}
which is a random variable with respect to $\mathbb{P}_{\lambda_{0}},\mathbb{P}^{\pi}$.  Then
\begin{equation}
\begin{array}{rl}
\mathbb{E}_{\mathbb{P}_{\lambda_{0}},\mathbb{P}^{\pi}} (B^{w}) \ = &
	M^{-1} \sum_{h=1}^{M}\mathbb{E}_{\mathbb{P}_{\lambda_{0}}} \mathbb{E}_{\mathbb{P}^{\pi}}\left[ \mathbbm{1} \left\{h \in S_{c}\right\} w_{h}a_{h0} \left( \dfrac{\hat{N}_{h}}
	{\hat{N}_{h} + \tau_{a}/\tau_{\epsilon}} \right)  \right.\\
&	+ \  \left.  \mathbbm{1} \left\{h \in S_{c}\right\} w_{h} \left( \dfrac{\sum_{j\in S_{h}} w_{j|h}\epsilon_{j h0}}
	{\hat{N}_{h} + \tau_{a}/\tau_{\epsilon}} \right)
		\right]\\
= &
	M^{-1} \sum_{h=1}^{M}\mathbb{E}_{\mathbb{P}_{\lambda_{0}}} \left[ \mathbb{E}_{\mathbb{P}^{\pi}} \left( \mathbbm{1} \left\{h \in S_{c}\right\} w_{h}a_{h0}\right) \left( \dfrac{\hat{N}_{h}}
	{\hat{N}_{h} + \tau_{a}/\tau_{\epsilon}} \right)  \right.\\
&	+ \  \left.   \mathbb{E}_{\mathbb{P}^{\pi}} \left( \mathbbm{1} \left\{h \in S_{c}\right\} w_{h} \right) \left(\dfrac{\sum_{j \in S_{h}} w_{j|h}\epsilon_{j h0}}
	{\hat{N}_{h} + \tau_{a}/\tau_{\epsilon}}\right)
		\right]\\
= &
	M^{-1}\sum_{h=1}^{M} \mathbb{E}_{\mathbb{P}_{\lambda_{0}}} \left[ (a_{h0}) \left( \dfrac{\hat{N}_{h}}
	{\hat{N}_{h} + \tau_{a}/\tau_{\epsilon}} \right)  \right.\\
&	+ \ \left.  \dfrac{\sum_{j\in S_{h}} w_{j|h}(\epsilon_{j h0})}
	{\hat{N}_{h} + \tau_{a}/\tau_{\epsilon}}
		\right]\\
= &
	M^{-1}\sum_{h=1}^{M} \mathbb{E}_{\mathbb{P}_{\lambda_{0}}} \left( \dfrac{\sum_{j\in S_{h}} w_{j|h}\epsilon_{j h0}}
	{\hat{N}_{h} + \tau_{a}/\tau_{\epsilon}} \right)	
\end{array}
\end{equation}
The population independence between $a_{h0}$ and the conditional inclusion probabilities (and weights) for individuals within cluster let us apply the expectation $\mathbb{E}_{\mathbb{P}_{\lambda_{0}}}$ independently between stages. Since $\mathbb{E}_{\mathbb{P}_{\lambda_{0}}}(a_{h0}) = 0$, the first term is 0; however, the within cluster conditional inclusion probabilities may in general depend on the noise $\epsilon_{jh0}$, therefore a joint expectation for the remaining, second term is needed in general. Condition \ref{itm:sampling design}\ref{withingroup} directly addresses this last term.
As the sampling within-cluster sampling fraction, $f_{c} = n_{h}/N_{h} < 1$, increases $\hat{N}_{h} = \mathcal{O}(N_{h})$ such that it becomes independent of $\mathbb{P}_{\lambda_{0}}$, which allows it to be factored.  So, for a sufficiently large $f_{c}$ the remaining term equals $0$ in the case that the expectation with respect to $\mathbb{P}_{\lambda_{0}}$ of the weighted sum of within-cluster residuals equals $0$.  It will be true for balanced within group informative sampling designs and nearly so for weakly unbalanced within group informative sampling designs, even under a small $f_{c}$, but certainly not for highly unbalanced sampling designs under a small $f_{c}$ (For an expanded discussion on different unbalanced sampling designs, see the simulations in Section~\ref{sec:sims2}).

Next, we define a second summary measure, the posterior conditional variance $\mathbb{V}$, to be:
\begin{equation}\label{varas}
\begin{array}{rl}
\mathbb{V}\left(M^{-1} \sum_{k \in S_{c}} w_k a_{k}\middle\vert\{y_{jk}\}, \{w_{jk}\} , \{w_{k}\}, \tau_{\epsilon}, \tau_{a} \right)  \ = &
M^{-2} \sum _{k \in S_{c}} w_{k}^{2} \left(\tau_\epsilon \sum_{j \in S_{k}} w_{jk} + \tau_{a}  w_{k} \right)^{-1}\\
 \ = & M^{-2} \sum _{h \in U_{c}} R^{W}_{h},
\end{array}
\end{equation}
where $R_{h}^{w}$ is the expanded estimator from observed units within each sampled cluster to the  population of units within each cluster in the population, which we achieve by inserting inclusion indicators, $\delta$, in,
\[
\begin{array}{rl}
R_{h}^{w} \ = & \mathbbm{1} \left\{h \in S_{c}\right\} w_{h}^{2} \left(\tau_{\epsilon}  \sum_{\ell = 1}^{N_{h}} \mathbbm{1} \left\{(\ell,h) \in S\right\} w_{\ell h}^{2} +  \mathbbm{1} \left\{h \in S_{c}\right\} w_{h}\tau_{a} \right)^{-1}.\\
= & w_{h} \left(\tau_{\epsilon}  \sum_{\ell = 1}^{N_{h}} \mathbbm{1} \left\{\ell \in S \middle\vert h \in S_{c}\right\} w_{\ell|h} + \tau_{a} \right)^{-1}.

\end{array}
\]
The supremum value, $
\sup_{\mathbb{P}_{\lambda_{0}}, \mathbb{P}^{\pi}} (R_{h}^{w})  = \gamma_{1}(\tau_{\epsilon} + \tau_{a})^{-1},
$
since the sum of conditional probabilities of selection within clusters $\sum_{\ell = 1}^{N_{h}} \pi_{\ell|h} \ge 1 \implies \sum_{\ell = 1}^{N_{h}} \mathbbm{1} \left\{\ell \in S \middle\vert h \in S_{c}\right\} \ge 1$ (i.e. when a cluster is selected, at least one member of the cluster is also selected).  The inverse gamma prior distributions restrict $(\tau_{a},\tau_{\epsilon}) > 0$, strictly.

Then the joint expectation for the conditional variance is bounded:
\begin{equation}
\begin{array}{rl}
\mathbb{E}_{\mathbb{P}_{\lambda_{0}}, \mathbb{P}^{\pi}}
\left[\mathbb{V}  \left( M^{-1}\sum_{h=1}^{M} \delta_{h} w_{h} a_{h}\middle\vert\cdot \right) \right] \ = &
M^{-2} \sum _{h \in U_{c}} \mathbb{E}_{\mathbb{P}_{\lambda_{0}}, \mathbb{P}^{\pi}}
\left[ R^{w}_{h}\right] \\
\ \le &  M^{-2} \gamma_{1}\sum _{h \in U_{c}} (\tau_{\epsilon} + \tau_{a})^{-1}\\
\ = & M^{-1}\gamma_{1}(\tau_{\epsilon} + \tau_{a})^{-1} = \mathcal{O}(M^{-1}).
\end{array}
\end{equation}
Thus, the joint expectation of the posterior conditional variance $\mathbb{E}_{\mathbb{P}_{\lambda_{0}}, \mathbb{P}^{\pi}}
\left[\mathbb{V}  \left( M^{-1}\sum_{h=1}^{M} \delta_{h} w_{h} a_{h}\middle\vert\cdot \right) \right] \rightarrow~0$ when $M\rightarrow \infty$.

Let $\mathbb{P}(\cdot)$ denote the measure $\mathbb{P}\left(\cdot\middle\vert\{y_{\ell h}: (\ell,h)\in S\}, \{\delta_{\ell h}\} , \{w_{h}: h \in S_{c}\}, \tau_{\epsilon}, \tau_{a} \right)$ associated with the pseudo posterior full conditional distribution (estimator). Then using Chebyshev, for any arbitrary $\delta > 0$,
\begin{equation}
\begin{array}{rl}
\mathbb{P} \left( \abs{ M^{-1}\sum_{h \in U_{c}}\delta_{h}w_{h}a_{h} - M^{-1}\sum_{h \in U_{c}}a_{h0}} > \delta  \right)
\ \le & \dfrac{\mathbb{V}  \left( M^{-1} \sum_{h \in U_{c}}\delta_{h}  w_{h} a_{h}\right)}{\delta^2}
 \implies \\
\mathbb{E}_{\mathbb{P}_{\lambda_{0}}, \mathbb{P}^{\pi}}
\left[
\mathbb{P} \left( \abs{M^{-1}\sum_{h \in U_{c}}\delta_{h}w_{h}a_{h} - M^{-1}\sum_{h \in U_{c}}a_{h0} } > \delta \right)\right]
\ \le &
\dfrac{\mathbb{E}_{\mathbb{P}_{\lambda_{0}}, \mathbb{P}^{\pi}}
\left[\mathbb{V}  \left( M^{-1} \sum_{h \in U_{c}} \delta_{h} w_{h} a_{h} \right) \right]}{\delta^2}\\
 \ =  &\mathcal{O}(N^{-1})
\end{array}
\end{equation}
since the integration in the joint expectations respects the inequalities. It is straightforward to show
the posterior distributions for all three averages
\begin{enumerate*}[label=(\roman*)]
\item $M^{-1}\sum_{h \in U_{c}}\delta_{h}w_{h}a_{h}$,
\item $M^{-1}\sum_{h \in U_{c}}a_{h}$, and
\item $M^{-1}\sum_{h \in U_{c}}a_{h0}$
\end{enumerate*} are each consistent estimators of $0$ at the same rate $\mathcal{O}(M^{-1})$. Then for $M$ sufficiently large we can readily substitute one expression for another.

We note that since we have the special case of $\mathbb{P}(\cdot) \in [0,1]$, the above expectation with respect to $\mathbb{P}_{\lambda_{0}}, \mathbb{P}^{\pi}$ implies an $L_{1}$ result $\left( \int \int \abs{\mathbb{P}(\cdot) - 0}d\mathbb{P}^{\pi} d\mathbb{P}_{\lambda_{0}} \right) \rightarrow 0$ and is stronger than (implies) results for $L_{k}$ for all $k > 1$. We use a similar approach for the remaining full conditional distributions.

\section{Proof of Theorem \ref{thm1}}
\label{sec:proof-theor-refthm1}

\subsection{Precision parameter of the random effects}

Next, we show contraction for the full conditional pseudo posterior of the precision for the random cluster effects.
Using Equation \ref{eq:1},  the first summary measure (expected value) of the conditional posterior is
\[
\mathbb{E} \left (\tau_{a}^{-1}\middle\vert\{a_{k}\}, \{w_{k}\}  \right) =
\dfrac{\sum_{k = 1}^{m} w_{k} a_{k}^{2} + 2\beta_1}{\sum_{k = 1}^{m}  w_{k} + 2\alpha_1-2}
\approx \dfrac{\sum_{k = 1}^{m} w_{k} a_{k}^{2}}{\sum_{k = 1}^{m}  w_{k}}
=\frac{1}{\hat{M}}  \sum_{k = 1}^{m} w_{k} a_{k}^{2},
\]
where the constants $\beta_1$ and $\alpha_1$ are negligible, because $m = \mathcal{O}(M) = \mathcal{O}(N)$. For simplicity we can invoke Slutsky to proceed with $M$ instead of $\hat{M}$. As before, we augment this summary measure from the observed sample to the population by inserting $\delta$ to achieve,
\[
G^{W} = \sum_{h = 1}^{M} \mathbbm{1} \left\{ h \in S_{c} \right\} w_{h} a_{h}^{2},
\]
Then,
\begin{equation}\label{eq:taumean}
\begin{array}{rl}
\mathbb{E}_{\mathbb{G}^{y}_{\theta_{0}}, \mathbb{P}^{\pi}} (G^{W} )  &=  \mathbb{E}_{\mathbb{G}^{y}_{\theta_{0}}, \mathbb{P}^{\pi}}\left[\sum_{h = 1}^{M} \mathbbm{1} \left\{ h \in S_{c} \right\} w_{h} a_{h}^{2}\right]\\
&=  \mathbb{E}_{\mathbb{G}^{y}_{\theta_{0}}}\left[
\sum_{h = 1}^{M} \mathbb{E}_{\mathbb{P}^{\pi}}\left(\mathbbm{1} \left\{ h \in S_{c} \right\}\middle\vert \lambda_{0}\right) w_{h} a_{h}^{2}\right]\\
&= \mathbb{E}_{\mathbb{G}^{y}_{\theta_{0}}}\left[
\sum_{h = 1}^{M} a_{h}^{2}\right]\\
&= \mathbb{E}_{\mathbb{G}^{y}_{\theta_{0}}}\left[
M M^{-1} \sum_{h = 1}^{M} a_{h}\right]^{2}\\
&= \mathbb{E}_{\mathbb{G}^{y}_{\theta_{0}}}\left[M\mathbb{E}_{\mathbb{P}^{\pi}}\left( M^{-1}\sum_{h = 1}^{M} \delta_{h} w_{h} a_{h} \right)\right]^{2}\\
&\ \stackrel{L_{1}-\mathbb{P}_{\lambda_{0},\mathbb{P}^{\pi}}}{\longrightarrow} \mathbb{E}_{\mathbb{P}_{\theta_{0}}}\left[M\mathbb{E}_{\mathbb{P}^{\pi}}\left( M^{-1}\sum_{h = 1}^{M} a_{h0} \right)\right]^{2}\\
&= \mathbb{E}_{\mathbb{P}_{\theta_{0}}}\left[M M^{-1}\sum_{h = 1}^{M} a_{h0}\right]^{2}\\
&= \mathbb{E}_{\mathbb{P}_{\theta_{0}}}\left[\sum_{h = 1}^{M} a_{h0}^{2}\right]\\
&= M \tau_{a0}^{-1},
\end{array}
\end{equation}
where in the third to seventh equations in Equation~\ref{eq:taumean} we have used the independence assumption of groups, $k \in (1,\ldots,M)$, of Condition~\ref{itm:sampling design} and the independence of the $\{a_{h0}\}_{h}$ such that $\mathbb{E}_{P_{\theta_{0}}}(a_{h0}a_{h'0}) = 0$ for $h,h^{'} \in (1,\ldots,M)$. This independence allows bringing the square inside the brackets in the second to last equation of Equation~\ref{eq:taumean}. We replace $a_{h}$ with $a_{h0}$ in the sixth equation since we have earlier shown that $M^{-1} \sum_{h} \delta_{h} w_{h} a_{h}$ contracts onto $M^{-1}\sum_{h}  a_{h0}$ in $L_{1}-\mathbb{P}_{\lambda_{0}}, \mathbb{P}^{\pi}$ for $M$ sufficiently large.  We achieve, $\mathbb{E}_{\mathbb{G}^{y}_{\theta_{0}}, \mathbb{P}^{\pi}} \left(\mathbb{E} \left (\tau_{a}^{-1}\middle\vert\{a_{h}: h \in S_{c}\}, \{\delta_{k}\},\{w_{h}:h\in S_{c}\}\right)\right)\rightarrow \tau_{a0}^{-1} $
as the number of population clusters $M \rightarrow \infty$.

Using Equation \ref{eq:2}, the variance of the conditional posterior is
\[
\mathbb{V} \left (\tau_{a}^{-1}\middle\vert\{a_{k}\}, \{w_{k}\}  \right)
\ \le \dfrac{C_{1}\left[\sum_{k = 1}^{m} w_{k} a_{k}^{2}\right]^2}{\left[\sum_{k = 1}^{m}  w_{k}\right]^3}
=\dfrac{C_{1}\left[ \sum_{k = 1}^{m} w_{k} a_{k}^{2} \right]^{2}}{\hat{M}^3},
\]
where $C_{1} > 1$ denotes a bounded constant. We, again, augment the numerator from the sample to the population by inserting $\delta$,
\[
H^{W} = \left[\sum_{h = 1}^{M} \mathbbm{1} \left\{ h \in S_{c} \right\} w_{h} a_{h}^{2}\right]^{2},
\]
Then
\begin{equation} \label{eq:tauexpvar}
\begin{array}{rl}
\mathbb{E}_{\mathbb{G}^{y}_{\theta_{0}}, \mathbb{P}^{\pi}} (H^{W} ) \ = &
  \mathbb{E}_{\mathbb{G}^{y}_{\theta_{0}},\mathbb{P}^{\pi}}\left[ \left(\sum_{h = 1}^{M}\delta_{h} w_{h} a_{h}^{2}\right) \left(\sum_{h^{'} = 1}^{M}\delta_{h^{'}} w_{h^{'}} a_{h^{'}}^{2}\right)\right]\\
\ = & \mathbb{E}_{\mathbb{G}^{y}_{\theta_{0}}, \mathbb{P}^{\pi}}\left(\sum_{h = 1}^{M}\delta_{h}^{2} w_{h}^{2} a_{h}^{4}\right) +  \mathbb{E}_{\mathbb{G}^{y}_{\theta_{0}}, \mathbb{P}^{\pi}}\left(\sum_{h \neq h = 1}^{M}\delta_{h}\delta_{h^{'}} w_{h}w_{h^{'}} a_{h}^{2}a_{h^{'}}^{2}\right)\\
\ = & \sum_{h=1}^{M}\mathbb{E}_{\mathbb{G}^{y}_{\theta_{0}}}\left(w_{h}a_{h}^{4}\right) + \sum_{h \neq h = 1}^{M}\mathbb{E}_{\mathbb{G}^{y}_{\theta_{0}}}\left(a_{h}^{2}\right)\mathbb{E}_{\mathbb{P}_{\lambda_{0}}}\left(a_{h^{'}}^{2}\right)
\\
\ \le & \gamma_{1}\left[\sum_{h=1}^{M}\mathbb{E}_{\mathbb{G}^{y}_{\theta_{0}}}\left(a_{h}^{4}\right) + \sum_{h \neq h = 1}^{M}\mathbb{E}_{\mathbb{G}^{y}_{\theta_{0}}}\left(a_{h}^{2}\right)\mathbb{E}_{\mathbb{G}^{y}_{\theta_{0}}}\left(a_{h^{'}}^{2}\right)\right]\\
\ = & \gamma_{1} \mathbb{E}_{\mathbb{G}^{y}_{\theta_{0}}} \left[\sum_{h = 1}^{M}  a_{h}\right]^{4}\\
\ = & \gamma_{1} \mathbb{E}_{\mathbb{G}^{y}_{\theta_{0}}} \left[M M^{-1}\sum_{h = 1}^{M}  a_{h}\right]^{4}\\
\ = & \gamma_{1}\mathbb{E}_{\mathbb{G}^{y}_{\theta_{0}}}\left[M \mathbb{E}_{\mathbb{P}^{\pi}} \left(M^{-1}\sum_{h=1}^{M}\delta_{h}w_{h}a_{h}\right)\right]^{4}
\\
\ \stackrel{L_{1}-\mathbb{P}_{\lambda_{0},\mathbb{P}^{\pi}}}{\longrightarrow} & \gamma_{1}\mathbb{E}_{\mathbb{P}_{\theta_{0}}}\left[M \mathbb{E}_{\mathbb{P}^{\pi}} \left(M^{-1}\sum_{h=1}^{M}a_{h0}\right)\right]^{4}
\\
\ = & \gamma_{1}\mathbb{E}_{\mathbb{P}_{\theta_{0}}}\left[\sum_{h=1}^{M}a_{h0}\right]^{4}
\\
\ = & \gamma_{1}\left[\{\sum_{h=1}^{M}\mathbb{E}_{\mathbb{P}_{\theta_{0}}}\left(a_{h0}^{4}\right)\} + M^{2}(\tau_{a0}^{-1})^{2}\right]\\
\ = &
3 \gamma_{1} \sum_{h = 1}^{M}  \{\mathbb{V}_{\mathbb{P}_{\theta_{0}}}(a_{h0} )\}^{2} + \gamma_{1}M^{2}(\tau_{a0}^{-1})^{2}\\
\ =  &  3 \gamma_{1} M \left(\tau_{a0}^{-1} \right)^{2} + \gamma_{1}M^{2}(\tau_{a0}^{-1})^{2}\\
\ \le & 4 \gamma_{1} M^{2} \left(\tau_{a0}^{-1} \right)^{2}
\end{array}
\end{equation}
The second term of the third equation from the top of Equation~\ref{eq:tauexpvar} relies on independent sampling across clusters from Assumption~\ref{itm:sampling design} under $\mathbb{P}^{\pi}$ and the a posteriori independence of the $\{a_k\}$ under $\mathbb{G}^{y}_{\theta_{0}}$. The third term from the bottom of Equation~\ref{eq:tauexpvar} is a property of the $4^{th}$ central moment of the normal distribution $\mathbb{E}(z^4) = 3\sigma^4$ for $z \sim N(0, \sigma^2)$.
Then since $a_h \perp a_{h'}$, their covariance is 0, and
\begin{equation}
\mathbb{E}_{\mathbb{G}^{y}_{\theta_{0}}, \mathbb{P}^{\pi}}\left[ \mathbb{V} \left (\tau_{a}^{-1}\middle\vert\{a_{h}:h\in S_{c}\}, \{\delta_{h}\}, \{w_{h}:h\in S_{c}\}   \right) \right] \le
\dfrac{4 \gamma_{1} M^{2} \left(\tau_{a0}^{-1} \right)^{2}}{M^3 } = \dfrac{4 \gamma_{1} \left(\tau_{a0}^{-1} \right)^{2} }{M} = \mathcal{O}(M^{-1})
\end{equation}

\subsection{Precision parameter of the noise}

Using  Equation \ref{eq:1}, the first summary measure for $\tau_{\epsilon}$, the expected value of the conditional posterior is,
\[
\begin{array}{rl}
\mathbb{E}\left (\tau_{\epsilon}^{-1}\middle\vert\{y_{jk}\}, \{a_{k}\}, \{w_{jk}\} , \mu \right) \ = &
\dfrac{\sum_{k = 1}^{m} \sum_{j = 1}^{n_{k}} w_{jk}\epsilon_{jk0}^2  + 2b_2}{ \sum_{k = 1}^{m} \sum_{j = 1}^{n_{k}} w_{jk}  + 2\alpha_2-2} \\
\ \approx & \dfrac{\sum_{k = 1}^{m} \sum_{j = 1}^{n_{k}} w_{jk}\epsilon_{jk0}^2}{ \sum_{k = 1}^{m} \sum_{j = 1}^{n_{k}} w_{jk} }\\
 \ = & \frac{1}{\hat{N}}  \sum_{k = 1}^{m} \sum_{j = 1}^{n_{k}} w_{jk}\epsilon_{jk0}^2.
\end{array}
\]
Similar to the construction for $\tau^{-1}_{a}$, we can show for this summary measure of the conditional posterior
\newline
$\mathbb{E}_{\mathbb{P}_{\lambda_{0}}, \mathbb{P}^{\pi}} \left[\mathbb{E} \left (\tau_{\epsilon}^{-1}\middle\vert\{y_{\ell h}:(\ell,h) \in S\}, \{a_{h}: h \in S_{c}\}, \{\delta_{\ell h}\},\{w_{\ell h}:(\ell h) \in S\} , \mu  \right)  \right] \rightarrow \tau_{\epsilon0}^{-1} $
as the number of  clusters in the population,  $M \rightarrow \infty$ (by expanding the posterior summary measure from the observed sample to the population by inserting random inclusion indicators,~$\delta$).  \newline
Let $\mathbb{E} \left(\tau_{\epsilon}^{-1}\middle\vert\cdot \right) \equiv \mathbb{E} \left( \tau_{\epsilon}^{-1}\middle\vert\{y_{\ell h}:(\ell,h) \in S\}, \{a_{h}: h \in S_{c}\}, \{\delta_{\ell h}\},\{w_{\ell h}:(\ell h) \in S\} , \mu  \right)$.
We take the expectation of the expanded estimator with respect to the joint distribution of population generation and the taking of a sample with,
\begin{equation}\label{eq:taueexp}
\begin{array}{rl}
 \mathbb{E}_{\mathbb{P}_{\lambda_{0}},\mathbb{P}^{\pi}}\left[\mathbb{E} \left(\tau_{\epsilon}^{-1}\middle\vert\cdot \right) \right]  \ = & N^{-1} \mathbb{E}_{\mathbb{P}_{\lambda_{0}},\mathbb{P}^{\pi}}\left[\sum_{h=1}^{M}\sum_{\ell=1}^{N_{h}}\mathbbm{1}\{(\ell,h)\in S\} w_{\ell h} \epsilon_{\ell h0}^{2}\right] \\
 \ = & N^{-1} \mathbb{E}_{\mathbb{P}_{\lambda_{0}},\mathbb{P}^{\pi}}\left[\sum_{h=1}^{M}\sum_{\ell=1}^{N_{h}} \epsilon_{\ell h0}^{2}\right] \\
 \ = & N^{-1} N \tau_{\epsilon0}^{-1}\\
 \ = & \tau_{\epsilon0}^{-1}.
\end{array}
\end{equation}

Using Equation \ref{eq:2}, variance of the conditional posterior is
\[
\mathbb{V}\left (\tau_{\epsilon}^{-1}\middle\vert\{y_{jk}\}, \{a_{k}\}, \{w_{jk}\} , \mu \right)
\le \dfrac{C_{2}\left[\sum_{k = 1}^{m} \sum_{j = 1}^{n_{k}} w_{jk}\epsilon_{jk0}^2\right]^2}{\left[\sum_{k = 1}^{m} \sum_{j = 1}^{n_{k}} w_{jk}\right]^3}
=\dfrac{C_{2}\left[ \sum_{k = 1}^{m} \sum_{j = 1}^{n_{k}} w_{jk}\epsilon_{jk0}^2 \right]^{2}}{\hat{N}^3},
\]
where $C_{2} > 1$ is a bounded constant. \newline
Let $ \mathbb{V}\left[\tau_{\epsilon}^{-1}\middle\vert\cdot\right] \equiv \mathbb{V} \left [\tau_{\epsilon}^{-1}\middle\vert\{y_{\ell h}:(\ell,h)\in S\}, \{a_{h}: h \in S_{c}\}, \{\delta_{\ell h}\}, \{w_{\ell h}: (\ell,h)\in S\} , \mu  \right]$.
\begin{equation}
\begin{array}{rl}
   \mathbb{E}_{\mathbb{P}_{\lambda_{0}}, \mathbb{P}^{\pi}}\left[ \mathbb{V} \left (\tau_{\epsilon}^{-1}\middle\vert\cdot \right) \right] \  \leq  & N^{-3}C_{2}\mathbb{E}_{\mathbb{P}_{\lambda_{0}}, \mathbb{P}^{\pi}}\left[\sum_{h=1}^{M}\sum_{\ell=1}^{N_{h}}\delta_{\ell h}w_{\ell h}\epsilon_{\ell h0}^{2}\right]^{2}  \\
    \ \le & N^{-3}C_{2}\gamma\mathbb{E}_{\mathbb{P}_{\lambda_{0}}}\left[\sum_{h=1}^{M}\sum_{\ell=1}^{N_{h}}\epsilon_{\ell h0}\right]^{4} \\
    \ = & N^{-3}C_{2}\gamma\left[\{\sum_{h=1}^{M}\sum_{\ell=1}^{N_{h}}\mathbb{E}_{\mathbb{P}_{\lambda_{0}}}(\epsilon_{\ell h0}^{4})\} + N^{2}(\tau_{\epsilon0}^{-1} )^{2}\right]\\
    \ = & N^{-3}C_{2}\gamma\left[3\sum_{h=1}^{M}\sum_{\ell=1}^{N_{h}}\{\mathbb{V}_{\mathbb{P}_{\lambda_{0}}}(\epsilon_{\ell h0} )\}^{2} + N^{2}(\tau_{\epsilon0}^{-1} )^{2}\right]\\
    \ = & N^{-3}C_{2}\gamma\left[3N(\tau_{\epsilon0}^{-1} )^{2} + N^{2}(\tau_{\epsilon0}^{-1} )^{2}\right]\\
    \ \le & N^{-3}C_{2}\gamma\left[4N^{2}(\tau_{\epsilon0}^{-1} )^{2}\right]\\
    \ = & 4 N^{-1} C_{2}\gamma (\tau_{\epsilon0}^{-1} )^{2}\\
    \ = & \mathcal{O} (N^{-1})
\end{array}
\end{equation}

\subsection{Intercept}\label{sec:amu}

Lastly, we examine the full conditional pseudo posterior distribution of the intercept $\mu$. Equation \ref{prec-tau} implies that
\begin{equation}
\begin{array}{rl}
\left( \mu\middle\vert\{y_{jk}\}, \{a_{k}\}, \{w_{jk}\}, \tau_{\epsilon} \right) \ & \propto \left[
\prod_{k=1}^{m} \prod_{j=1}^{n_{k}} \mathcal{N}\left(\tilde{y}_{jk}\middle\vert\mu, \tau_{\epsilon}\right)^{w_{ij}}
\right] \times \mathbbm{1}\{\mu \in (-\infty, \infty)\}\\
\ & = \ N\left (\dfrac{\sum_{k = 1}^{m} \sum_{j = 1}^{n_{k}} w_{jk}\tilde{y}_{jk}}{\sum_{k = 1}^{m} \sum_{j = 1}^{n_{k}} w_{jk}}, \left[\tau_{\epsilon}\sum_{k = 1}^{m} \sum_{j = 1}^{n_{k}} w_{jk}\right]^{-1} \right)
\end{array}
\end{equation}
where $\tilde{y}_{jk} = y_{ij} - a_{k}$.

Proposition \ref{prop1} implies that $M^{-1}\sum_{h = 1}^{M} \delta_{h} w_{h} a_{h} \rightarrow M^{-1}\sum_{h = 1}^{M} a_{h0}$ as $M \rightarrow \infty$ in $L_{1}-\mathbb{P}_{\lambda_{0}}, \mathbb{P}^{\pi}$. Therefore, we may replace $a_h$ with $a_{h0}$ in $\delta_{\ell h}\tilde{y}_{\ell h} = \delta_{\ell h}(\mu + \epsilon_{\ell h0})$ to form $\sum_{h=1}^{M}\sum_{\ell=1}^{N_{h}}\delta_{\ell h}w_{\ell h}(\mu + \epsilon_{\ell h0})$.  Then, using similar population constructions as above, we can show that
\[\mathbb{E}_{\mathbb{P}_{\lambda_{0}}, \mathbb{P}^{\pi}}\left[\mathbb{E} (\mu\middle\vert\{y_{\ell h}:(\ell,h)\in S\}, \{a_{h}:h\in S_{c}\}, \{\delta_{\ell h}\}, \{w_{\ell h}:(\ell,h)\in S\}, \tau_{\epsilon}  ) \right] = \mu_{0} \] and that
\[\mathbb{E}_{\mathbb{P}_{\lambda_{0}}, \mathbb{P}^{\pi}}\left[\mathbb{V}^{-1} (\mu\middle\vert\{y_{\ell h}:(\ell,h)\in S\}, \{a_{h}:h\in S_{c}\}, \{\delta_{\ell h}\}, \{w_{\ell h}:(\ell,h)\in S\}, \tau_{\epsilon}  ) \right] = \tau_{\epsilon} N \] and thus $\mathbb{E}_{\mathbb{P}_{\lambda_{0}}, \mathbb{P}^{\pi}}\left[\mathbb{V} (\mu\middle\vert\cdot ) \right] = \mathcal{O} (N^{-1})$.

\section{Simulation Study 3: Sensitivity to Diminishing Cluster Sampling Fraction $\frac{m}{M} \rightarrow 0$}\label{sec:sims3}
As noted in Section \ref{sec:assume}, we assume cluster sampling fraction $\frac{m}{M}$ that is bounded above 0, assumption~\ref{itm:bounds}\ref{constantfrac}. In a related assumption, we assume that the minimum cluster inclusion probability $\pi_{h}$ is bounded away from 0 and the corresponding weight $w_{h} < \gamma_1$ is bounded above by some constant. We now demonstrate a pathological case, in which the sample fraction $\frac{m}{M} \rightarrow 0$ in such a way that some $\pi_{h} \rightarrow 0$ faster than others, with rates related to $a_{h}$, resulting in biased estimates to illustrate that our double weighted method is inconsistent under these conditions as asserted by \citet{slud_2019}. 

In practice, however, we would expect most practical designs to have similar rates for $\pi_{h} \rightarrow 0$ across $h$, which would be $\mathcal{O}\left(\frac{m}{M}\right)$. When we enforce this by using a design protected against this pathological case, the double weighted method exhibits fairly robust estimation and near consistency.

\subsection{Sample Design and Estimation}\label{sec:sim3des}
The model is the same as in Section \ref{sec:sims1model}. The sample design is similar to that of Section \ref{sec:sim2des}, however we modify the PPS size measures for the first stage and increase the population  clusters $M$ as a faster rate than the sample clusters $m$:

\begin{enumerate}
\item Generate R = 100 populations from the one-way ANOVA distribution (using true values, $\mu_{0} = 1$, $\tau_{a}^{-1} = 2$, $\tau_{\epsilon}^{-1} = 3$).
	\begin{enumerate}
		\item With $M = $ \{4,000; 16,000; 64,000\} clusters
		\item Each with $N_{h} = \{40\}$ individuals in each cluster
	\end{enumerate}
\item For each $r = 1,\ldots, R$ population, draw a two-stage sample via `mstage'  in R \citep{sampling}
	\begin{enumerate}
		\item Sample $m = \{20,40,60\}$ clusters. Using systematic PPS sampling with size:
		\begin{enumerate}
			\item  $\pi_h \propto \max(0,a_{h}) + 1/M$
		\end{enumerate}
		\item Sample $n_{k} = \{5\}$ individuals in each cluster design. Using systematic PPS sampling with size
		\begin{enumerate}
			\item  $\pi_{\ell|h} \propto \epsilon_{h \ell} - \min_{h \ell}+ 1$
		\end{enumerate}
	\end{enumerate}
\item For each $r = 1,\ldots, R$ sample, estimate $\{b_0, \sigma_a, \sigma_{\epsilon}\}$
	\begin{enumerate}
		\item Using equal weights $w_{k} = 1$,  $w_{k j} = 1$ via `lmer' in R \citep{lme4}
		\item Using double weights $w_{k} \propto 1/\pi_{k}$,  $w_{k j} \propto 1/\pi_{k j}$ via Stan \citep{stan:2015}.
	\end{enumerate}
\end{enumerate}
The size measure $\pi_h \propto \max(0,a_{h}) + 1/M$ has the effect of splitting the population of clusters in half like two strata. For $a_h < 0$, the effective strata are taken as an SRS with faster diminishing sampling rate of $\mathcal{O}\left(\frac{m}{M^2}\right)$. For $a_h > 0$, the clusters are sampled PPS, with sample fraction still of rate $\mathcal{O}\left(\frac{m}{M}\right)$.

\subsection{Results}
\begin{figure}
\centering
\includegraphics[width = 0.95\textwidth,
		page = 1,clip = true, trim = 0.0in 0.0in 0in 0.in]{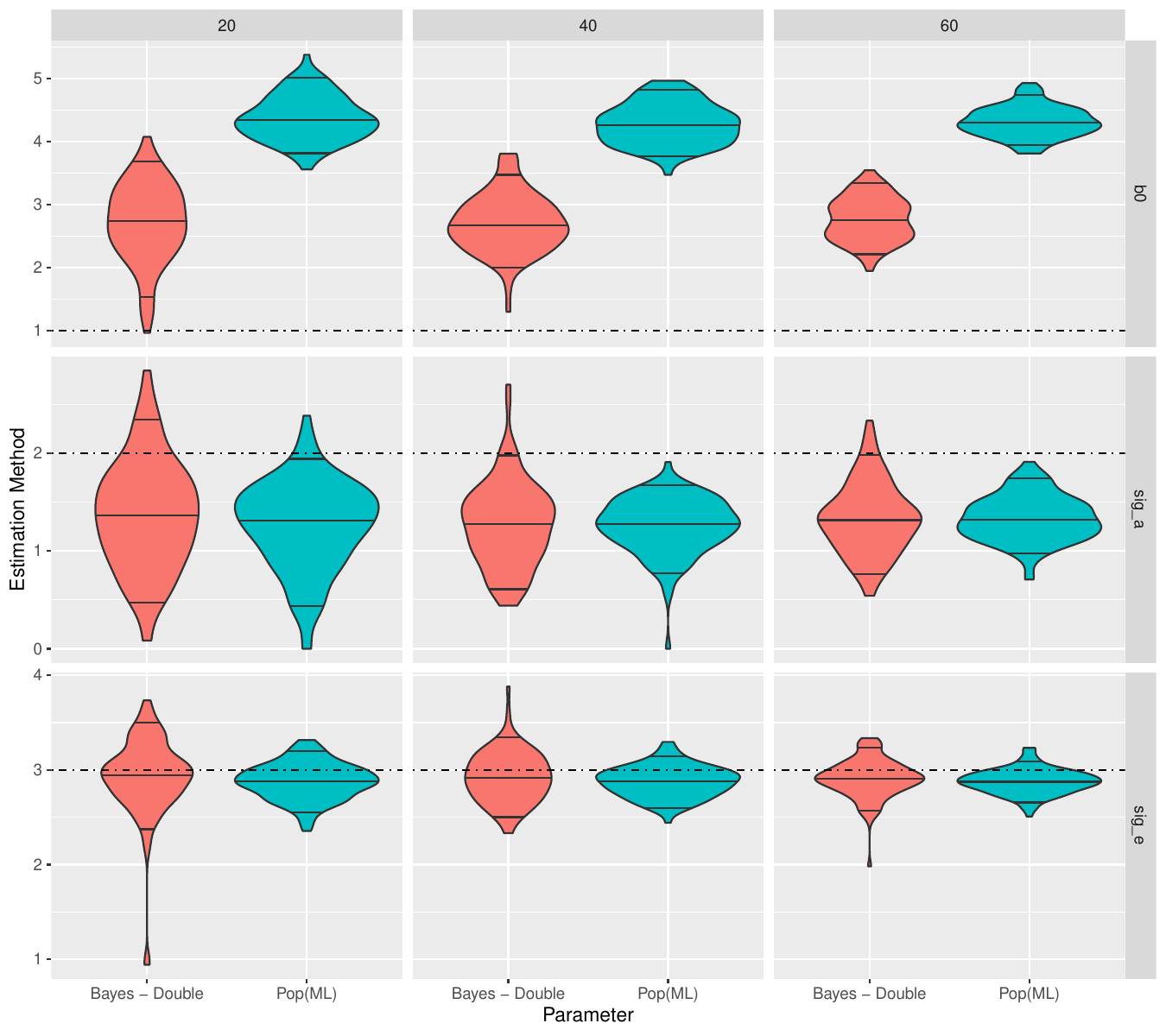}
\caption{Distributions and quantiles (5\%, 50\%, 95\%) of parameter estimates for R = 100 simulations for estimators (x-axis) and parameters (rows) across cluster sample sizes (cols) for \emph{pathological} first stage sample design with linear asymmetric second stage sample design. Reference lines: population generating values.}
\label{fig:dw_asymfc}
\end{figure}

Figure~\ref{fig:dw_asymfc} presents the Monte Carlo simulation distributions for each of the parameters, $(\mu,\tau_{a}^{-1},\tau_{\epsilon}^{-1})$, in the rows, from top-to-bottom.  The left-hand plots in each plot panel are our double-weighted estimator while the accompanying right-hand plots are for an unweighted MLE that doesn't correct for informativeness, as a comparison.  

We can readily see that our double weighted method is now \emph{inconsistent} for the case of that cluster sampling fraction $m/M$ limits to $0$ \emph{and} where the sampling design is constructed such that $\pi_{h} \downarrow 0$ at $\mathcal{O}(\frac{m}{M^{2}})$ for some clusters (such that the associated $w_{h}$ limits to $\infty$) in our two strata cluster design.  While such a sampling design is possible, it is not typical; for example, sampling designs used at the Bureau of Labor Statistics will bound all inclusion probabilities strictly away from $0$.  

While not shown here, we conducted additional simulation studies under more typical sampling designs represented by the linear asymmetric cluster sampling design of Section~\ref{sec:sim2des} where $\pi_{h}$ decreases linearly at $\mathcal{O}(\frac{m}{M})$ (by linearly reducing inclusion probabilities to achieve the target $m$) and let  $m/M \downarrow 0$.  We achieve the same nearly consistent result under the linear sampling design for the double weighted method as shown in Figure~\ref{fig:compare_asymclust} under the linear asymmetric design, even as we let $m/M \downarrow 0$.

All to say, even when our assumption that the cluster sampling fraction is bounded away from $0$ is violated, our double weighted estimator is still generally consistent so long as Assumption~\ref{itm:sampling design}\ref{withingroup} is met and the decrease in $\pi_{h}$ is limited to a linear, $\mathcal{O}(\frac{m}{M})$ rate.   It is only in a corner case that $\pi_{h}$ decline at a non-linear $\mathcal{O}(\frac{m}{M^2})$ rate for some $h \in (1,\ldots,M)$ that our double weighted estimator becomes inconsistent as predicted by \citet{slud_2019}.

\end{document}